\documentclass[twocolumn,showpacs,preprintnumbers,amsmath,amssymb]{revtex4}
\usepackage{indentfirst}
\usepackage{hhline}
\usepackage{wrapfig, float}
\usepackage{array,longtable,tabularx}
\usepackage{graphicx}% Include figure files
\usepackage{dcolumn}% Align table columns on decimal point
\usepackage{bm}% bold math
\usepackage{subfigure}
\usepackage[small]{caption2}
\usepackage{boxhandler}
\usepackage{color}
\usepackage{hyperref}
\usepackage{textcomp}% qu
\renewcommand{\captionlabeldelim}{.}
\hypersetup{linkcolor=blue, citecolor=blue, urlcolor=blue, colorlinks=true}%, pdfborderstyle={/S 1}}
\begin{document}

%\begin{center}

%\preprint{1507.08513}

\title{Lattice instability and enhancement of superconductivity in YB$_6$}

\author{N.~Sluchanko}
\thanks{corresponding author; e-mail: nes@lt.gpi.ru}
\affiliation{Prokhorov General Physics Institute of RAS, 38 Vavilov Street, 119991 Moscow, Russia}
\affiliation{Moscow Institute of Physics and Technology, 9 Institutskiy Lane, 141700 Dolgoprudny, Moscow Region, Russia}\author{A.~Azarevich}\affiliation{Prokhorov General Physics Institute of RAS, 38 Vavilov Street, 119991 Moscow, Russia}\author{M.~Anisimov}\affiliation{Prokhorov General Physics Institute of RAS, 38 Vavilov Street, 119991 Moscow, Russia}\author{A.~Bogach}\affiliation{Prokhorov General Physics Institute of RAS, 38 Vavilov Street, 119991 Moscow, Russia}\author{S.~Gavrilkin}\affiliation{Lebedev Physical Institute of RAS, 53 Leninskiy Avenue, 119991 Moscow, Russia}\author{V.~Glushkov}\affiliation{Prokhorov General Physics Institute of RAS, 38 Vavilov Street, 119991 Moscow, Russia}\affiliation{Moscow Institute of Physics and Technology, 9 Institutskiy Lane, 141700 Dolgoprudny, Moscow Region, Russia}\author{S.~Demishev}\affiliation{Prokhorov General Physics Institute of RAS, 38 Vavilov Street, 119991 Moscow, Russia}\affiliation{Moscow Institute of Physics and Technology, 9 Institutskiy Lane, 141700 Dolgoprudny, Moscow Region, Russia}\author{K.~Mitsen}\affiliation{Lebedev Physical Institute of RAS, 53 Leninskiy Avenue, 119991 Moscow, Russia}\author{A.~Kuznetsov}\affiliation{National Research Nuclear University MEPhI, 31 Kashirskoe Shosse, 115409 Moscow, Russia}\author{I.~Sannikov}\affiliation{National Research Nuclear University MEPhI, 31 Kashirskoe Shosse, 115409 Moscow, Russia}\author{N.~Shitsevalova}\affiliation{Frantsevich Institute for Problems of Materials Science of NASU, \\ 3 Krzhizhanovskogo Street, 03680 Kiev, Ukraine}
\author{V.~Filippov}\affiliation{Frantsevich Institute for Problems of Materials
Science of NASU, \\ 3 Krzhizhanovskogo Street, 03680 Kiev, Ukraine}
\author{M.~Kondrin}\affiliation{Vereshchagin Institute for High Pressure Physics of RAS, 142190 Troitsk, Russia}
\author{K.~Flachbart}\affiliation{Institute of Experimental Physics of
SAS, 47 Watsonova Street, SK-04001 Ko\v{s}ice, Slovak Republic}

\date{\today}
\begin{abstract}
The superconducting
and normal state characteristics of yttrium hexaboride (YB$_6$) have been investigated for the
single crystals with a transition temperatures $T_c$ ranging between 6~K and 7.6~K.
The extracted set of microscopic parameters [the coherence length $\xi$(0) $\sim$ 320$\div$340~${\textmd{\AA}}$,
the penetration depth $\lambda$(0) $\sim$ 1100$\div$1600~${\textmd{\AA}}$ and the mean free path of charge carriers \textit{l} = 31$\div$58~${\textmd{\AA}}$, the Ginzburg-Landau-Maki parameters $\kappa$$_{\textmd{1,2}}$(0) $\sim$ 3.3$\div$4.8 and the superconducting
gap $\Delta$(0) $\sim$ 10.3$\div$14.8~K] confirms the type II superconductivity in
"dirty limit" ($\xi$$\gg$\textit{l}) with a medium to strong electron-phonon interaction (the electron-phonon interaction constant
$\lambda_{\textmd{e-ph}}$ = 0.93$\div$0.96) and
\textit{s}-type pairing of charge carriers in this compound [2$\Delta$(0)$/k_BT_c$ $\approx$ 4].
The comparative analysis of charge transport (resistivity, Hall and Seebeck coefficients)
and thermodynamic (heat capacity, magnetization) properties in the normal
state in YB$_6$ allowed to detect a transition into the cage-glass state
at $T^*$ $\sim$ 50~K with a static disorder in the arrangement of the Y$^{3+}$ ions. We argue that the significant
$T_c$ variations in the YB$_6$ single crystals are determined by two main factors: (\textit{i}) the superconductivity enhancement is related with the increase of the number of isolated vacancies, both at yttrium and boron sites, which leads to the
development of an instability in the hexaboride lattice; (\textit{ii}) the $T_c$
depression is additionally stimulated by the spin polarization of conduction electrons emerged and enhanced by the magnetic field in the
vicinity of defect complexes in the YB$_6$ matrix.
\end{abstract}

\maketitle

%\textbf{1.}
\section*{I. INTRODUCTION}

The discovery of high-temperature superconductivity in magnesium diboride (MgB$_2$) with $T_c$ $\sim$ 39~K \cite{1} stimulated an interest in the study of microscopic mechanisms, which are responsible for the appearance of superconductivity in higher borides $R$B$_6$ and $R$B$_{12}$ with a rigid covalent framework composed of boron clusters. The maximum transition temperature in these model superconductors $-$ cage-glasses \cite{2} was observed in yttrium hexaboride YB$_6$ ($T_c$ $\sim$ 8~K \cite{3}) in which the pairing  is mainly influenced by low-energy ($\sim$ 8~meV) Einstein-like quasi-local vibrations of loosely bound yttrium ions located in the B$_{24}$ cubooctahedra of the boron sub-lattice.

The strong $T_c$ dispersion (1.5$\div$8.4~K) reported for YB$_6$ samples  by different authors \cite{3}-\cite{8} has no satisfactory explanation up to now. For example, the YB$_6$ single crystals grown by a modified Al-Ga flux growth method in Al$_2$O$_3$ crucible under Ar pressure show $T_c$ = 5.8$\pm$0.1~K regardless on the initial composition. The $T_c$ values for samples grown by argon arc melting are 6.8$\div$7.0~K. For YB$_6$ with nominal compositions obtained by ultrafast quenching from melt the starting point of the resistive transition to superconducting state was observed at $T_c$ $\sim$ 8.4~K \cite{4}. Superconducting transition temperatures of YB$_6$ single crystals grown by induction zone melting also reach high values $T_c$ = 7.2~K \cite{3}, 7.5~K \cite{5,6} and 7.7~K \cite{7}. The lowest values of $T_c$ = 1.5$\div$6.3~K were reported for powder samples prepared by the borothermal reduction \cite{8}. To explane such a significant $T_c$ variation it was suggested in \cite{3} that the transition temperature is controlled by the B/Y ratio (the highest $T_c$ was obtained for a B/Y$<$6). Thus, both a growth of a number of boron vacancies, which is associated with the deviation from the stoichiometric composition of the boron sub-lattice, and a decrease of yttrium vacancies in contrast, which requires a stoichiometric metal sub-lattice composition, result according to \cite{3} into a $T_c$ enhancement in this compound. However, this conclusion contradicts clearly to observations made in \cite{6} where the highest $T_c$ was observed on sample with the lowest residual resistivity (i.e., with the lowest number of defects). Moreover, whereas the microanalysis results \cite{3} indicate a significant concentration of boron vacancies (a YB$_{5.7 \div 5.8}$ composition), the most YB$_6$ single crystals grown by zone melting \cite{9} correspond approximately to a composition YB$_{6.1}$, which points to the presence of yttrium vacancies in the hexaboride matrix. It should be emphasized that some of the mentioned methods of YB$_6$ synthesis correspond to non-equilibrium crystallization conditions. This fact allows to assume that the superconducting transition temperature can be significantly modified in the non-equilibrium, metastable state of yttrium hexaboride. A fairly large residual resistivity $\rho_0$ $\sim$ 8$\div$25~$\mu\Omega$ cm and a rather small residual resistivity ratio  $\rho$(300K)/$\rho_0$ = 2$\div$4.5 observed for all the YB$_6$ single crystals studied up to now also indicates a strong low temperature scattering of charge carriers on crystal structure defects and inhomogeneities, which can be associated also with a non-equilibrium state of yttrium hexaboride.

Superconducting $T_c$ enhancement in the vicinity of lattice instabilities in the non-equilibrium state is a well-known effect which is up to now not well understood in detail. For example, the amorphous beryllium films deposited by evaporation on low temperature substrates show $T_c$ of about 10~K enhanced if compared with $T_c$ = 0.026~K for the \textit{hcp} phase of Be \cite{10,11}. The value of $T_c$ in Ga thin films prepared by condensation at low temperatures increases up to 8.4~K from that of 1.1~K for bulk gallium \cite{12}. Non-equilibrium Al$_{1-x}$Si$_x$ solid solutions demonstrate a $T_c$ variation between 1.18~K (\textit{x} = 0) and 11~K (\textit{x} $\sim$ 0.2), their superconductivity enhancement being attributed to a lattice instability developed in these \textit{fcc} Al--based crystals \cite{13}$-$\cite{16}. Therefore, it is interesting to consider YB$_6$ crystals with different $T_c$ values from the point of view of non-equilibrium superconductivity establishing relationship between the degree of the \textit{bcc} lattice instability and $T_c$ changes.

In this context the presented results of detailed studies of specific heat, magnetization, resistivity, Hall and Seebeck coefficients and hydrostatic density in YB$_6$ single crystals with different values of $T_c$ in the range between 6~K and 7.6~K allowed us to elucidate the mechanism, which is responsible for the enhancement/suppression of superconductivity. We argue that the lattice instability, which develops in YB$_6$ under an increase of the number of isolated vacancies on boron and yttrium sites of the boride matrix, is the main factor controlled the observed $T_c$ enhancement. On the contrary, the local accumulation of single vacancies into complexes produces strong distortions of the YB$_6$ lattice, which result in a suppression of superconductivity in this compound.

The paper is organized as follows: Experimental details and results are shown in sections \hyperref[Sec.2]{II} and \hyperref[Sec.3]{III}, respectively. In the discussion part \hyperref[Sec.4p1]{IV.1} the data analysis of the superconducting state is presented and we argue in favor of type II superconductivity in the "dirty limit" with a medium to strong electron-phonon interaction and \textit{s}-type pairing of charge carriers in YB$_6$. In part \hyperref[Sec.4p2]{IV.2} a detailed analysis of the normal state parameters is undertaken which allowed us to conclude that below $T^*$ $\sim$ 50~K a cage-glass state forms in YB$_6$. Final conclusions are formulated in Section \hyperref[Sec.5]{V}.

%%%% ----------------------------------------------------------------------
\section*{II. EXPERIMENTAL DETAILS  }\label{Sec.2}

\begin{figure}[t]
\begin{center}
\includegraphics[width = 8.7cm]{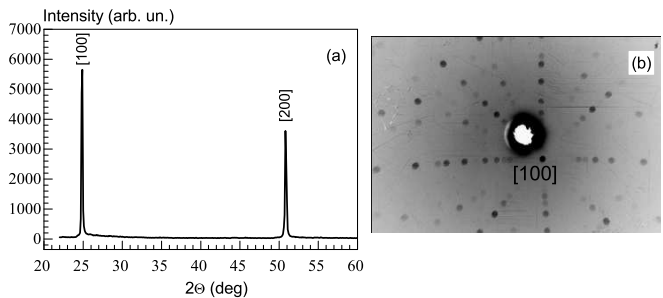}
   \parbox{8cm}{\caption{The X-ray diffraction pattern of the polished plate cut perpendicular to the growth axis obtained in the Co K$_\alpha$ radiation with a Fe filter [panel (a)] and Laue backpattern [panel (b)] in Co K$_a$ radiation for crystal YB$_6$ No.2 ($T_c$ = 7.4~K).}}\label{FigX1}
   \end{center}
\end{figure}

Measurements were performed on three single crystals of yttrium hexaboride with $T_c$ = 6.6~K (No.3), 7.4~K (No.2) and 7.55~K (No.1) [according to results of zero-field resistivity measurements]. The studied single crystals were grown by induction zone melting in IPM NASU, Kiev, using rods sintered from powder obtained by borothermal reaction of yttrium oxide (Y$_2$O$_3$) with a purity of 99.999~$\%$ and amorphous boron having a purity of $>$99.5~$\%$. Taking into account the nature of the peritectic melting of YB$_6$, we synthesized the initial powder with boron excess. Crystal growth from the boron enriched melt allowed to (\textit{i}) decrease the melting temperature below the peritectic one (2600~C), (\textit{ii}) obtain single-phase ingots and (\textit{iii}) improve the real structure of crystals. The optimum boron composition of the initial sintered rods was consistent with YB$_{6.65}$$-$YB$_{6.85}$. Other optimization parameters were the pressure of high purity argon gas in the growth chamber (0.7~MPa) and the crystallization rate (0.22~mm/min). Because of the zone cleaning effect during the process of crystal growth the impurity concentration did not exceed 0.001~wt.$\%$. In order to control the composition of samples we used additional optical emission spectral and microanalysis techniques. The quality and single-phase of crystals were controlled by X-ray methods. As an example, Fig.\hyperref[FigX1]{1} shows the diffraction pattern of a polished plate cut perpendicularly to the growth axis [panel \hyperref[FigX1]{(a)}] and the Laue backscattering pattern [panel \hyperref[FigX1]{(b)}] of one YB$_6$ crystal (No.2, with $T_c$ = 7.4~K). The obtained value of the lattice constant $a$ = 4.1001$\pm$0.0005~${\textmd{\AA}}$ is identical within the experimental accuracy for all three investigated single crystals. The heat capacity and Hall effect were measured using a Quantum Design installation PPMS-9 in the Shared Facility Centre of Lebedev Physical Institute of RAS in the temperature range 1.9--300~K and in magnetic fields up to 9~T. Field and temperature dependences of magnetization were recorded both by a Quantum Design MPMS-5 and a SQUID magnetometer \cite{17}. For measurements of resistivity and thermoelectric power we used the original setup described in \cite{18,19}, respectively. The technique applied for the measurement of hydrostatic density of samples is described in detail in \cite{20}.

\section*{III.  RESULTS AND DISCUSSION}\label{Sec.3}
\subsection*{\emph{3.1. Resistivity.}}

\begin{figure}[t]
\begin{center}
\includegraphics[width = 8.2cm]{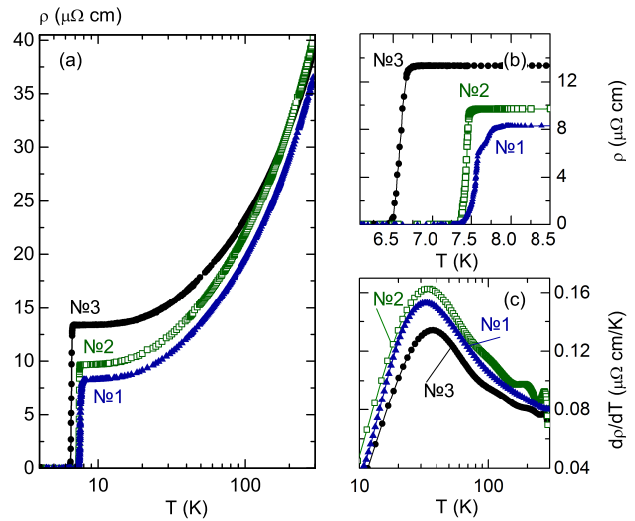}
  \parbox{8cm}{ \caption{(Colour on-line) (a-b) Temperature dependences of resistivity $\rho$(\textit{T}) for different YB$_6$ samples. Panel (b) shows the region of the superconducting transition. (c) Temperature dependence of the resistivity derivative $d\rho$($T$)/$dT$.}}\label{FigX2}
   \end{center}
\end{figure}

Figure \hyperref[FigX2]{2(a)} shows the temperature dependences of resistivity  $\rho$($T$) of all three studied YB$_6$ crystals. The $\rho$($T$) curves exhibit a typical metallic behavior with a rather small residual resistivity ratio $\rho$(300K)/$\rho_0$ = 2.9$\div$4.5. The residual resistivity of sample No.1 with the highest $T_c$ = 7.55~K is the smallest ($\rho_0$ $\sim$ 8~$\mu\Omega$~cm). The increase of $\rho_0$ is accompanied by a decrease of $T_c$ [Fig.\hyperref[FigX2]{2(a)}] which is in accordance with results of \cite{6} and \cite{3}. Figure \hyperref[FigX2]{2(b)} shows the $\rho$($T$) dependence in the vicinity of the superconducting transition. For all YB$_6$  single crystals studied we observed a wide enough resistivity transition with a width of  $\Delta$$T_c^{(\rho)}$ $\sim$ 0.12$\div$0.3~K [Table \hyperref[Tab.1]{I}] as well as non-monotonous $\rho$($T$) behavior near $T_c$, which is a particularly discerned for sample No.1 with maximal $T_c$. The $T_c^{(\rho)}$ values found as mid-points $\rho$($T_c$) = 1/2$\rho_0$ of resistivity transitions are shown in Table \hyperref[Tab.1]{I}. Fig.\hyperref[FigX2]{2(c)} demonstrates the corresponding temperature derivatives $d\rho/dT$.

\subsection*{\emph{3.2. Specific heat.}}

\begin{figure}[t]
\begin{center}
\includegraphics[width = 7.9cm]{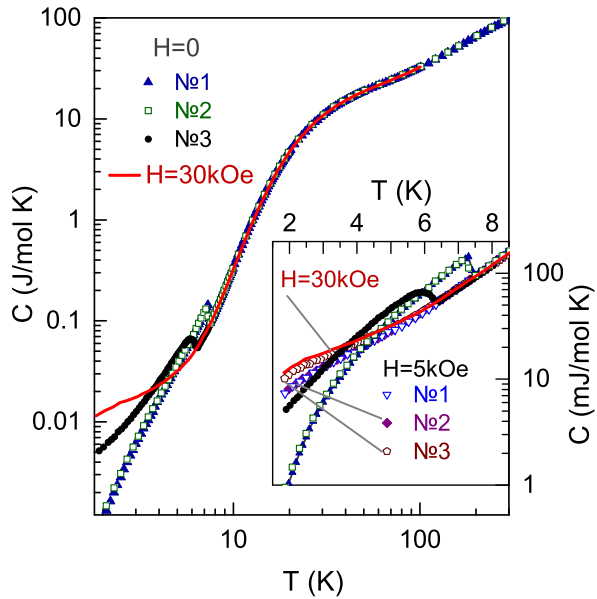}
   \parbox{8cm}{ \caption{(Colour on-line) Temperature dependences of the heat capacity for different YB$_6$ samples measured in zero magnetic field and at $H$ = 30~kOe [sample No.3]. The inset shows an enlarged area around the transition temperature; data at $H$ = 5~kOe correspond to the normal state of YB$_6$.}}\label{FigX3}
   \end{center}
\end{figure}

The heat capacity temperature dependences $C$($T$) of the investigated YB$_6$ single crystals are shown in Fig.\hyperref[FigX3]{3}. The inset in Fig.\hyperref[FigX3]{3} highlights the heat capacity behavior in the vicinity of the superconducting phase transition. In addition, Figure \hyperref[FigX3]{3} shows also the $C$($T$) curves measured in magnetic fields of 5 and 30~kOe in which the superconductivity of yttrium hexaboride is completely supressed. As can be seen in Figure \hyperref[FigX3]{3}, a gradual diminution of the heat capacity at temperatures between 300~K and 50~K is followed by a sharp almost step-like decrease with a typical Einstein-type $C$($T$) dependence below 40~K. It is worth noting that in the normal state at $H \geq$ 5~kOe and $T >$ 6~K the $C$($T$) curves of all three YB$_6$ samples are almost identical in the double logarithmic plot used in Figure \hyperref[FigX3]{3}. At the same time, there is an essential difference in the heat capacity of samples No.1, No.2 on one side, and sample No.3, on the other side, observed both in the superconducting and normal state at lowest temperatures [see inset in Figure \hyperref[FigX3]{3}].

\begin{figure}[b]
\begin{center}
\includegraphics[width = 8.4cm]{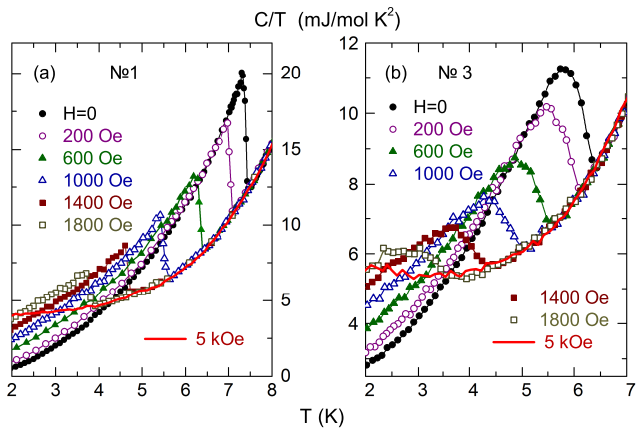}
   \parbox{8cm}{ \caption{(Colour on-line) Dependences of the low temperature heat capacity of YB$_6$ samples (a) No.1 and (b) No.3, measured in different external magnetic fields $H \leq $ 5~kOe.
 }}\label{FigX4}
   \end{center}
\end{figure}

\begin{figure}[t]
\begin{center}
\includegraphics[width = 8.4cm]{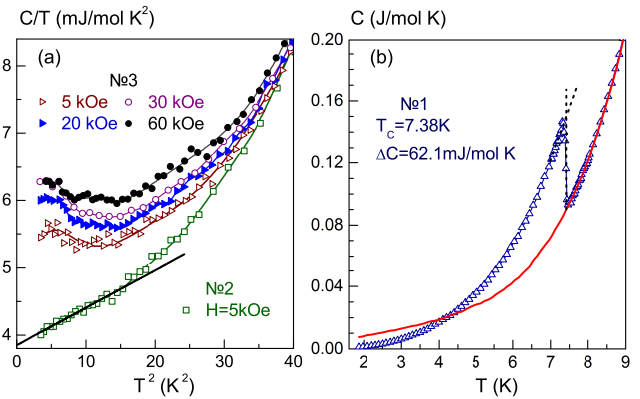}
   \parbox{8cm}{ \caption{(Colour on-line) (a) Dependences of the low temperature heat capacity of YB$_6$ in the coordinates $C/T$ vs. $T^2$ for samples No.2 and No.3 [panel (a)] for different values of external magnetic field $H \leq$ 60~kOe. Panel (b) shows the procedure applied to determine the $\Delta C$ jump amplitude near $T_c$ in sample No.1.
 }}\label{FigX5}
   \end{center}
\end{figure}

The results of specific heat measurements at low temperatures and in small magnetic fields which just destroy superconductivity are presented in Figure \hyperref[FigX4]{4}. For comparison, heat capacity curves are shown in this figure for samples No.1 and No.3 with a significantly different $T_c$ [see panels \hyperref[FigX4]{(a)} and \hyperref[FigX4]{(b)}, respectively] in coordinates $C$($T$, $H_0$)/$T$ vs. $T$. Apart from $T_c$ changes between samples No.1, No.2 and No.3, there are also differences related with both lowering of the jump amplitude $\Delta$$C$ near $T_c$ and with the broadening of this anomaly [see panel \hyperref[FigX4]{(b)} in Figure \hyperref[FigX4]{4} and Table \hyperref[Tab.1]{I}]. Moreover, it is necessary to note the marked increase of the absolute values of the specific heat of sample No.3, which can be observed in superconducting and normal state at temperatures $T <$ 6~K. Figure \hyperref[FigX5]{5(a)} shows the low-temperature heat capacity of samples No.2 and No.3 in coordinates $C$($T$, $H_0$)/$T$  vs. $T^2$, which is commonly used to determine the Sommerfeld coefficient $\gamma$ of the electronic heat capacity. For YB$_6$ crystals No.1 and No.2 the obtained values  $\gamma$ = 3.8$\div$3.85~mJ/(mol K$^2$) coincide with each other, whereas the low temperature specific heat of sample No.3  is obviously influenced by a moderate additional magnetic contribution. It should be mentioned here that magnetic defects, clusters and spin glass behavior can result into a specific heat enhancement \cite{21}, and lead in some cases also to a false indication of heavy fermion behavior \cite{22,23}. In such cases ivestigations of magnetic field changes of the heat capacity can help to identify the nature of the enhancement. For this reason we have carried out field dependent heat capacity measurements on crystal No.3. The received magnetic component demonstrates a moderate increase in external magnetic field [Fig.\hyperref[FigX5]{5(a)}], but the results do not allow to estimate properly the value of the electronic contribution.

In Fig.\hyperref[FigX5]{5(b)} the determination procedure of the transition temperature $T_c$ and of the heat capacity jump at $T_c$ are shown for sample No.1. The magnitude of the jump $\Delta$$C$ was determined as the length of the vertical line between the asymptotics of temperature dependences of the specific heat in the normal and superconducting states [see Fig.\hyperref[FigX5]{5(b)}]. The obtained $\Delta$$C$ values are presented in Table \hyperref[Tab.1]{I}. Note that the superconducting transition temperatures deduced from heat capacity measurements are smaller ($\Delta$$T_c$ $\sim$ 0.15$\div$0.3~K), than those obtained both from resistivity [Figure \hyperref[FigX2]{2(a)}] and field-cooled ($H$ $\sim$ 4$-$8~Oe) magnetization curves [see inset in Figure \hyperref[FigX6]{6} and Tables \hyperref[Tab.1]{I} and \hyperref[Tab.2]{II} for the comparison of $T_c$ values]. To minimize the errors of $T_c$ evaluation a special calibration procedure of the temperature sensors used in PPMS-9
(Quantum Design) and  in the

%_________________________________________________________________________________________
\begin{longtable}[t]{ccccc}%\firsthline
\caption{Superconducting state parameters obtained from heat capacity measurements: $T_c^{(C)}$ and  $\Delta$$T_c^{(C)}$ are the transition temperature and the width of transition, $H_{cm}$ and $H_{c2}$ the thermodynamic and the second critical fields, $\Delta C$ the heat capacity jump at $T_c$,  $\Delta$(0) the superconducting gap, $\kappa_1$(0) the Ginzburg-Landau-Maki parameter, $\xi$(0) the coherence length and $\lambda$(0) the penetration depth. Also shown are the transition temperature $T_c^{(\rho)}$ and the width of the transition  $\Delta$$T_c^{(\rho)}$, obtained from resistivity measurements.}\label{Tab.1} \\
\\
\hhline{=====}\\
\quad\quad & \quad\quad & \quad No.1 \quad\quad\quad & \quad No.2 \quad\quad\quad & \quad No.3 \quad\quad\\
\\
\hline\\
$T_c^{(\rho)}$/$T_c^{(C)}$ (K) & \quad\quad & 7.55/7.38 & 7.4/7.3  & 6.6/6.2  \\
 & \quad\quad & & & \\
$\Delta$$T_c^{(\rho)}$/$\Delta$$T_c^{(C)}$ (K) &\quad\quad & 0.3/0.15 & 0.12/0.15 & 0.2/0.4 \\
 & \quad\quad & & & \\
$H_{cm}$~(Oe) & \quad\quad & 618 & 613 & 429 \\
 & \quad\quad & & & \\
$\Delta$$C$~(mJ/mol K) & \quad\quad & 62.1 & 59 & 29.3 \\
 & \quad\quad & & & \\
$\Delta C$/$\gamma T_c$ &\quad\quad & 2.21 & 2.1 & 1.24 \\
 & \quad\quad & & & \\
$\Delta$(0)~(K) &\quad\quad & 14.8 & 14.6 & 12.1 \\
 & \quad\quad & & & \\
2$\Delta$(0)/$T_c$ &\quad\quad & 4.01 & 3.99 & 3.91 \\
 & \quad\quad & & & \\
$H_{c2}$(0)~(Oe) &\quad\quad & 2850 & 2912 & 2927 \\
 & \quad\quad & & & \\
$d H_{c2}$/$dT$~(Oe/K) &\quad\quad & --559 & --575 & --623 \\
 & \quad\quad & & & \\
$\kappa_1$(0) &\quad\quad & 3.26 & 3.36 & 4.82 \\
& \quad\quad & & & \\
$\xi$(0)~(${\textmd{\AA}}$) &\quad\quad & 340 & 336.4 & 335.5 \\
& \quad\quad & & & \\
$\lambda$(0)~(${\textmd{\AA}}$) &\quad\quad & 1109 & 1130 & 1618 \\ %\lasthline
\\
\hhline{=====} \\
\end{longtable}
%__________________________________________________________________________________________
\noindent
installation for resistivity measurements \cite{10} was carried out. The obtained differences between the $T_c$ data from resistivity [$T_c^{(\rho)}$] and heat capacity measurements [$T_c^{(C)}$] are therefore probably caused due to the presence of very small number of phases with high $T_c$ values, which within the experimental accuracy could not be detected in heat capacity measurements. As a result, we will consider the $T_c$ values obtained from heat capacity and magnetization measurements at $H$ $\sim$ 20~Oe as the characteristics of bulk superconductivity in the studied YB$_6$ samples.

\subsection*{\emph{3.3. Magnetization.}}

\begin{figure}[t]
\begin{center}
\includegraphics[width = 7.0cm]{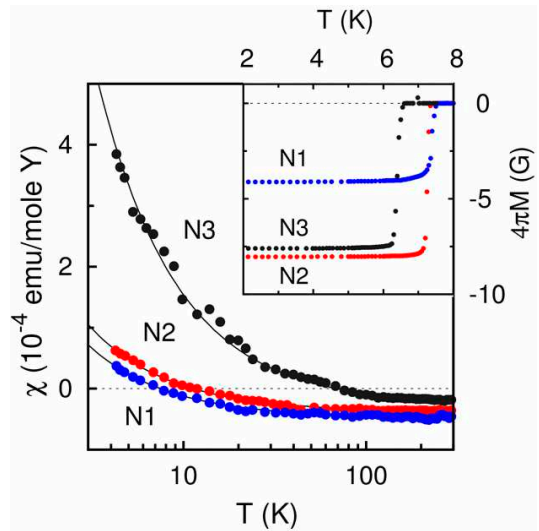}
    \parbox{8cm}{\caption{(Colour on-line) Temperature dependences of the magnetic susceptibility $\chi$(\textit{T}) = $M$($T$)/$H$ for different samples of YB$_6$, measured in magnetic field of 5.4~kOe. Solid lines show the fitting of experimental data by Eq.\hyperref[Eq.16]{(16)}. The inset shows the superconducting transition measured during cooling at $H$ = 4~Oe (sample No.1), 8~Oe (No.2) and 7.5~Oe (No.3).
}}\label{FigX6}
   \end{center}
\end{figure}

Figure \hyperref[FigX6]{6} shows the temperature dependence of magnetic susceptibility $\chi$($T$) = $M$($T$)/$H$ of the YB$_6$ samples as deduced from magnetization measured at $H_0$ = 5.4~kOe. It is visible from Fig.\hyperref[FigX6]{6}, that in the normal state of YB$_6$ the susceptibility significantly increases with temperature lowering, changing from negative values at $T >$ 100~K to positive ones at low temperatures. As a result, the presence of two additive components in the normal state should be taken into account $-$ the paramagnetic contribution caused by localized magnetic moments of magnetic impurities and the diamagnetic component originating from the YB$_6$ matrix. It can be discerned in Figure \hyperref[FigX6]{6} that the low-temperature component of the paramagnetic susceptibility of sample No.3 exceeds significantly [$\sim$ 5 times] the $\chi$($H$, $T$) values of No.1 and No.2 crystals.

\begin{figure*}[t]
\begin{center}
\includegraphics[width = 11.5cm]{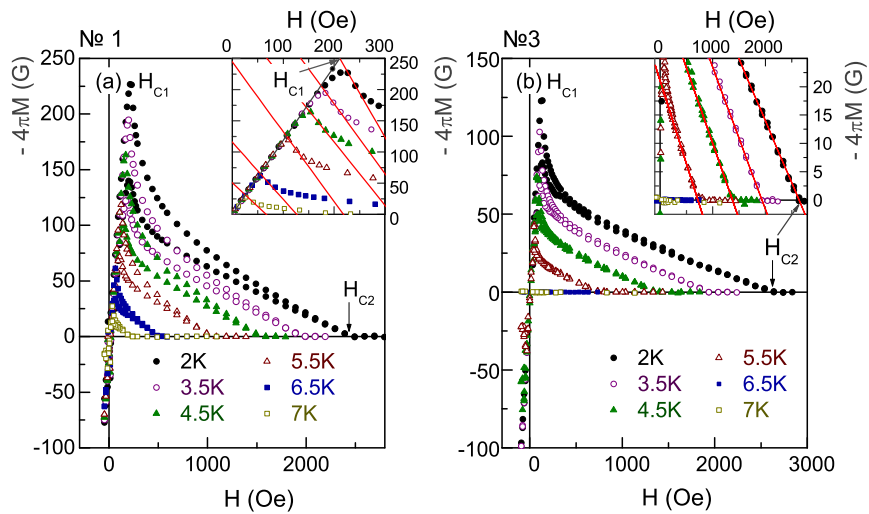}
\parbox{17cm}{\caption{(Colour on-line) Magnetic field dependences of magnetization $M$($H$, $T_0$) in the superconducting state and in the vicinity of the transition  temperature for samples (a) No.1 and (b) No.3. The insets show the procedure for determining the critical fields $H_{c1}$ and $H_{c2}$ (see also text). }}\label{FigX7}
   \end{center}
\end{figure*}

\begin{figure}[b]
\begin{center}
\includegraphics[width = 8.4cm]{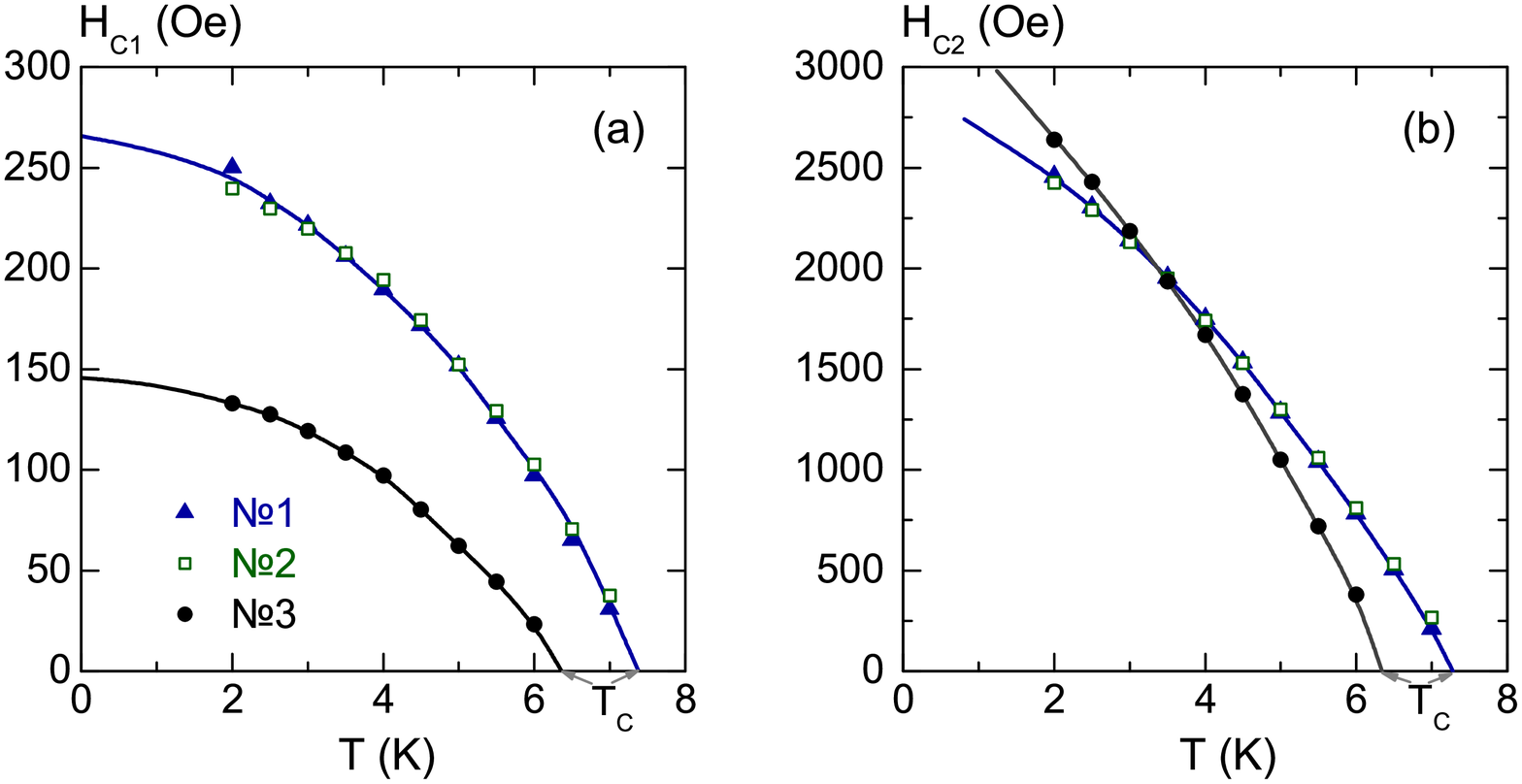}
   \parbox{8cm}{\caption{(Colour on-line) Temperature dependences of (a) the lower $H_{c1}$ and (b) the upper $H_{c2}$ critical fields for different YB$_6$ samples resulting from magnetization measurements. }}\label{FigX8}
   \end{center}
\end{figure}

Below the transition temperature $T_c$ a diamagnetic response is observed on magnetization curves $M$($T$) in small magnetic fields, and this superconducting component corresponds within experimental accuracy to the total Meissner effect [see inset in Figure \hyperref[FigX6]{6}]. An increase of external magnetic field up to 3~kOe leads to the appearance of features on $M$($H$, $T_0$) curves which are typical for type II superconductors. Indeed, a linear rise of the diamagnetic magnetization is observed in the range below the first critical field $H < H_{c1}$ corresponding to Meissner phase, and above $H_{c1}$, in the mixed state, $M$($H$) decreases dramatically until the transition at the second critical field $H_{c2}$ to the normal state occurs. Fig.\hyperref[FigX7]{7} demonstrates the diamagnetic $M$($H$, $T_0$) dependences as obtained for samples No.1 and No.3 [panels \hyperref[FigX7]{(a)} and \hyperref[FigX7]{(b)}, respectively]. The procedure usually applied for the extraction of critical fields is shown in the insets of Fig.\hyperref[FigX7]{7}, where the intersection points of linear asymptotics marked as $H_{c1}$ and $H_{c2}$ are shown for various temperatures. The values of $H_{c1}$ were corrected to the demagnetization factor which varies between 1.05 [sample No.3] and 1.185 [sample No.1]. The received behavior of $H_{c1}$($T$) and $H_{c2}$($T$) for all three studied YB$_6$ crystals is presented in Fig.\hyperref[FigX8]{8}. It can be seen that the critical fields $H_{c1}$($T$) and $H_{c2}$($T$) for samples No.1 and No.2 almost coincide with each other [Fig.\hyperref[FigX8]{8}], while a much smaller  $H_{c1}$($T$) and both higher $dH_{c2}/dT$ values of the derivative at $T_c$ and $H$$_{c2}$(0) correspond to sample No.3.

\subsection*{\emph{3.4. Hall and Seebeck coefficients.}}

\begin{figure*}[t]
\begin{center}
\includegraphics[width = 11.5cm]{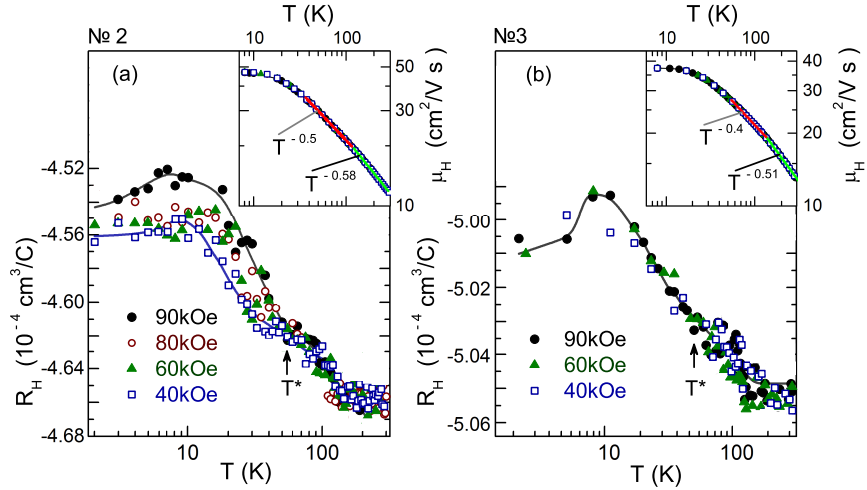}
   \parbox{17cm}{\caption{(Colour on-line) Temperature dependences of the Hall coefficient $R_H$($T$) obtained for samples (a) No.2 and (b) No.3 of YB$_6$ at different magnetic fields 40$\div$90~kOe. The arrows at $T^*$ indicate the phase transition to the cage-glass state. The insets show the temperature dependence of the Hall mobility $\mu_H$($T$) = $R_H$($T$)/$\rho$($T$). The solid lines in the insets demonstrate the approximation of the Hall mobility by power law dependence $\mu_H$($T$) $\sim$ $T^{-\alpha}$ (see text). }}\label{FigX9}
   \end{center}
\end{figure*}

For samples No.2 and No.3 the results obtained from Hall resistivity and thermopower measurements are shown in Figs.\hyperref[FigX9]{9} and \hyperref[FigX10]{10}, respectively; the data are plotted as Hall and Seebeck coefficients $R_H$($T$) and $S$($T$). As can be seen in Fig.\hyperref[FigX9]{9}, the Hall coefficient of YB$_6$ is negative and its magnitude slightly decreases with decreasing temperature in the range 2$\div$300~K. The absolute value of $R_H$ corresponds to a carrier concentration $n_e/n_\textmd{Y}$ = 0.9$\div$0.95 which is slightly below one electron per yttrium ion. Only moderate changes of $R_H$($T$) are observed with the temperature  lowering beeing in the limit of 2.2~$\%$ and 1.4~$\%$ for crystals  No.2 and No.3, respectively. The most significant decrease of $R_H$($T$) is detected in the range between 10 and 50~K corresponding to the step-like anomaly in the temperature dependence of heat capacity [see Fig.\hyperref[FigX3]{3}]. Within the limits of experimental accuracy, as the related studies were performed only in fields between 40 and 90~kOe, the data of Fig.\hyperref[FigX9]{9} do not allow to discuss the dependence of the Hall coefficient on external magnetic field.

The temperature dependences of the Seebeck coefficient [Fig.\hyperref[FigX10]{10}] demonstrate a typical metallic behavior $-$ the magnitude of $S$($T$) changes from negative values $\sim$ 1$\div$3~$\mu$V/K at intermediate temperatures 80$\div$300~K to small alternating ones $-$0.5$\div$0.5~$\mu$V/K in the low temperature range. As a result, two main features of thermopower can be detected: (\textit{i}) a peak near $T_c$ and (\textit{ii}) a $S$($T$) maximum in the vicinity of $T^*$ $\sim$ 50~K [see Fig.\hyperref[FigX10]{10}] which corresponds to previously discussed anomaly of the Hall coefficient [Fig.\hyperref[FigX9]{9}]. In the range between these two features a minimum on $S$($T$) curves is observed. Then, in the superconducting state, the thermopower decreases sharply to close to zero values [inset in Fig.\hyperref[FigX10]{10}] which are typical for superconductors \cite{24}.

\section*{IV. DISCUSSION}\label{Sec.4}
\subsection*{\emph{4.1. Characteristics of the superconducting state of YB$_6$.}}\label{Sec.4p1}
\subsubsection*{\emph{4.1.1 Analysis of specific heat.}}

\begin{figure}[t]
\begin{center}
\includegraphics[width = 8.9cm]{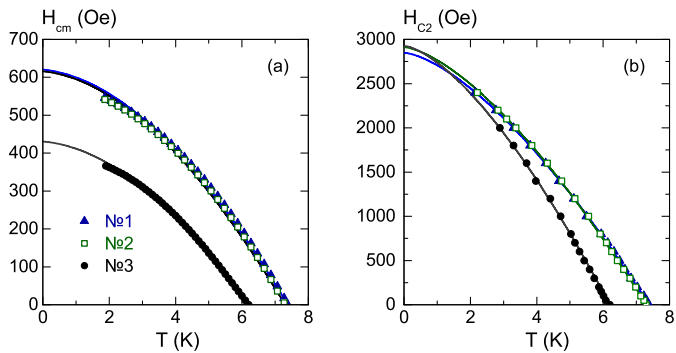}
   \parbox{8cm}{\caption{(Colour on-line) Temperature dependences of (a) the thermodynamic $H_{cm}$ and (b) the upper $H_{c2}$ critical fields for different YB$_6$ samples resulting from specific heat measurements. The solid lines show the data approximation by Eqs.\hyperref[Eq.4]{(4)} and \hyperref[Eq.5]{(5)}.}}\label{FigX11}
   \end{center}
\end{figure}

The specific heat results in the normal and superconducting states [Figs.\hyperref[FigX3]{3}--\hyperref[FigX4]{4}] were used to determine the thermodynamic critical field $H_{cm}$($T$) within the framework of standard relations
\begin{equation}\label{Eq.1}
-\frac{1}{2} \mu_0 V H_{cm}^2 (T) = \Delta F(T) = \Delta U(T)- T \Delta S(T)
\end{equation}
\begin{equation}\label{Eq.2}
 \Delta U(T)= \int_T^{T_c} [C_s(T')- C_n(T')]dT'
\end{equation}
\begin{equation}\label{Eq.3}
\Delta S(T) = \int_T^{T_c} \frac{[C_s(T')- C_n(T')]}{T'} dT' ,
\end{equation}
\begin{figure}[b]
\begin{center}
\includegraphics[width = 7cm]{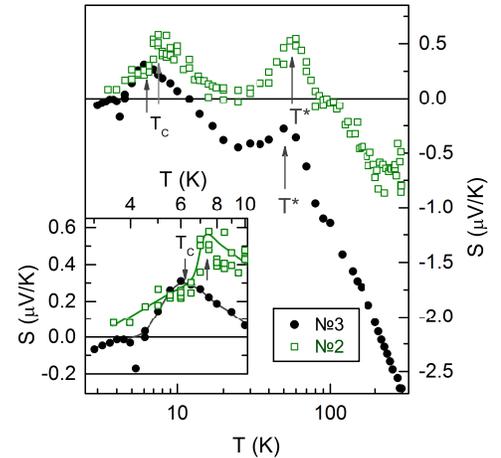}
   \parbox{8cm}{\caption{(Colour on-line) Temperature dependences of the Seebeck coefficient $S$($T$) for samples No.2 and No.3. The inset shows an enlarged area near the superconducting transition temperature $T_c$. The arrows indicate $T_c$ and the transition to the cage-glass phase ($T^*$). }}\label{FigX10}
   \end{center}
\end{figure}
where $F$ and $U$ denote the free and internal energies, $S$ -- the entropy, $V$ -- the molar volume, and the indices $n$ and $s$ correspond to characteristics of the normal and superconducting phases of YB$_6$. The integration was carried out in the temperature range from $T$ to $T_c$. Before the integration the specific heat data in the normal and superconducting states were approximated by polynomials of the 4-th order. Figs.\hyperref[FigX11]{11(a)} and \hyperref[FigX11]{11(b)} show the dependences of the thermodynamic $H_{cm}$($T$) and upper critical fields $H_{c2}$($T$), respectively, resulting from the heat capacity analysis of studied crystals. Table \hyperref[Tab.1]{I} presents the $H_{cm}$(0) values obtained by extrapolation of $H_{cm}$($T$) curves in the framework of the standard Bardeen-Cooper-Schriffer (BCS) relation
\begin{eqnarray}\label{Eq.4}
H_{cm} (T)/H_{cm} (0) = 1.7367 (1 - T/T_c) \nonumber\\
\times[1 - 0.327 (1 - T/T_c) - 0.0949(1 - T/T_c)^2].
\end{eqnarray}
In addition, Table \hyperref[Tab.1]{I} presents also the derivatives $dH_{c2}/dT$ at $T = T_c$ obtained from experimental data and the upper critical field $H_{c2}$(0) defined within the framework of formula used in \cite{25}
\begin{equation}\label{Eq.5}
H_{c2} (0) = -0.69 T_c \left(\frac{dH_{c2}}{dT}\right)_{T = T_c} .
\end{equation}
Using the value of the electronic specific heat coefficient  $\gamma$  = 3.8$\div$3.85~mJ/(mol K$^2$) received for crystals No.1 and No.2 [see Figure \hyperref[FigX5]{5}], the density of electronic states at the Fermi level $N_b$($E_F$) = 0.119 (eV atom)$^{-1}$ known from band structure calculations \cite{26,27} and the relation $\gamma$  = 1/3$\pi^2$$k_B^2$$N_b$($E_F$)(1 + $\lambda_{e-ph}$) [k$_B$ -- Boltzmann constant], we estimate the electron-phonon interaction constatnt $\lambda_{e-ph}$ = 0.93$\div$0.96 -- in good agreement with results of \cite{3}.

\begin{figure}[t]
\begin{center}
\includegraphics[width = 5.6 cm]{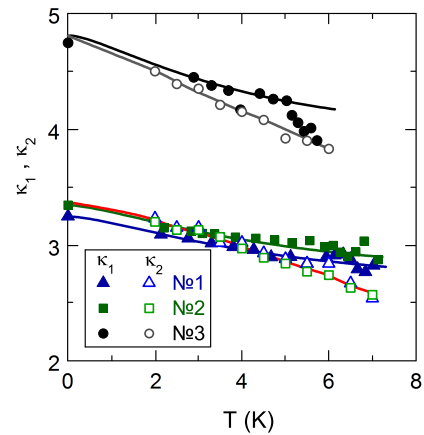}
   \parbox{8cm}{\caption{(Colour on-line) Temperature dependence of the Ginzburg-Landau-Maki parameters obtained from measurements of heat capacity ($\kappa_1$) and magnetization ($\kappa_2$) of different YB$_6$ samples.}}\label{FigX12}
   \end{center}
\end{figure}

Then, from BCS relations
\begin{equation}\label{Eq.6}
\Delta (0) = \frac{H_{cm}(0)}{\sqrt{2 \pi N(E_F)}}
\end{equation}
\begin{equation}\label{Eq.7}
\xi (0) = \sqrt{ \frac{\Phi_0}{2 \pi H_{c2}}}
\end{equation}
\begin{equation}\label{Eq.8}
\kappa_1 (T) = \frac{H_{c2} (T)}{\sqrt{2} H_{cm} (T)},
\end{equation}
where $\Phi_0$ denotes the flux quantum, the Ginzburg-Landau-Maki parameter $\kappa_1$($T$) \cite{28} (Fig.\hyperref[FigX12]{12}), the superconducting gap $\Delta$(0), the coherence length $\xi$(0) and the penetration depth $\lambda$(0) = $\kappa_1$(0)/$\xi$(0) [Table \hyperref[Tab.1]{I}] could be calculated. For sample No.3, due to the presence of a magnetic contribution to heat capacity [see Fig.\hyperref[FigX5]{5}] and the associated problem with the determination of the Sommerfeld coefficient, the estimation of $\gamma$ was obtained from the relation  $\gamma$$T_c^2$/$H_{cm}^2$(0) = \textit{const} \cite{3}. The ratio of 2$\Delta$/$k_B T_c$ $\approx$ 3.9$\div$4 found in this study for all studied YB$_6$ samples coincides with results obtained both in \cite{3}, with the heat capacity analysis of \cite{29, 30} and with the point-contact and tunnel spectra of \cite{29, 30}, and it significantly exceeds the value of 3.52 of the BCS model. It is also worth noting that the smaller values of 2$\Delta$/$k_B T_c$ $\approx$ 3.8 found in \cite{31} and 3.6 in \cite{32} from ultra-high resolution photoemission spectra at 5~K and tunneling spectra at 4.3~K, respectively, obviously may be attributed to gap $\Delta$($T$) lowering at about 5~K [close to $T_c$ $\approx$ 7~K] in comparison with $\Delta$(0).

\subsubsection*{\emph{4.1.2 Analysis of magnetization.}}

The analysis of magnetization was carried out based on formulas which are well-known from the Abrikosov theory of type-II superconductivity \cite{33}
\begin{equation}\label{Eq.9}
- 4 \pi M = (H_{c2} - H)/[(2 \kappa_2^2 - 1) \beta_{\Delta}]
\end{equation}
\begin{equation}\label{Eq.10}
H_{c1}(T) = H_{c2}/2 \kappa_2^2 (ln \kappa_2 + a),
\end{equation}
where  $\kappa_2$ is the Ginzburg-Landau-Maki parameter \cite{28,34}, $\beta_{\Delta}$ = 1.16 the coefficient corresponding to a triangular lattice of Abrikosov vortices, and $a$ the constant depending on impurity concentration. Presented in Fig.\hyperref[FigX7]{7} are the linear dependences of magnetization $M$($H$) in the superconducting phase near $H_{c2}$ which allow to derive the $\kappa_2$($T$) behavior within the framework of Eq.\hyperref[Eq.9]{(9)} [see Fig.\hyperref[FigX12]{12}]. Then, the extrapolation to zero temperature provides the values of $\kappa_2$(0) and $a$ parameters. In addition, the use of relation \hyperref[Eq.7]{(7)} allows to estimate the coherence length $\xi$(0) and the penetration depth  $\lambda$(0) = $\kappa_2$(0)/$\xi$(0) [see Table \hyperref[Tab.2]{II}].

The comparison of Ginzburg-Landau-Maki parameters $\kappa_1$($T$) and  $\kappa_2$($T$) \cite{28,33,34} for crystals No.1 -- No.3 obtained from the analysis of heat capacity [Eq.\hyperref[Eq.8]{(8)}] and magnetization [Eq.\hyperref[Eq.9]{(9)}], respectively, shows that $\kappa_1$ and $\kappa_2$ differ mainly near $T_c$, but their characteristics are practically identical at temperatures below $T_c$/2 [Fig.\hyperref[FigX12]{12}] 
resulting to the relation  $\kappa_1 \geq \kappa_2$ for temperatures below $T_c$. However, according to \cite{34,35} in case of a superconductor in the "dirty" limit an opposite inequality $\kappa_1 \leq \kappa_2$ is expected for any relation between the mean free path $l$ of charge carriers and the coherence length $\xi$. Our estimates of $l$ from residual resistivity  $\rho_0$, from the Hall coefficient $R_H$, and from parameters $\xi$(0) and $\Delta$(0) lead within the framework of standard relations
\begin{equation}\label{Eq.11}
l = R_H  m^* v_F/(e \rho_0)
\end{equation}
\begin{equation}\label{Eq.12}
\xi (0) = \hbar v_F/[ \pi \Delta(0)]
\end{equation}
\begin{figure}[t]
\begin{center}
\includegraphics[width = 8.3cm]{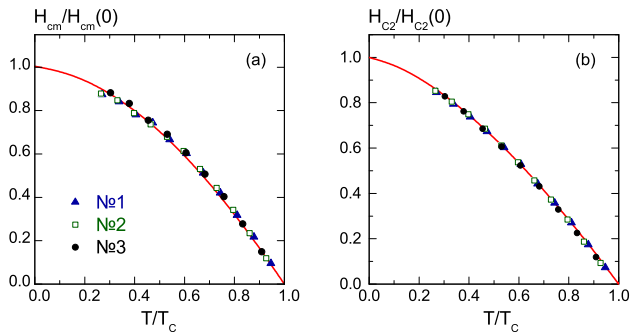}
   \parbox{8cm}{\caption{(Colour on-line) Scaling of the results obtained from magnetic measurements: (a) the normalized thermodynamic ratio $H_{cm}$/$H_{cm}$(0) and (b) the upper critical $H_{c2}$/$H_{c2}$(0) fields plotted as a function of reduced temperature for various samples.}}\label{FigX13}
   \end{center}
\end{figure}
\noindent
(where $v_F$ is the average Fermi velocity and $m^*$ the effective mass, $m^*$ = 1.03$m_0$ \cite{36}) to values of \textit{l} = 31$\div$58~${\textmd{\AA}}$ for the studied crystals [see Table \hyperref[Tab.3]{III}]. This results also to inequality $l \ll \xi$ that validates the "dirty limit" for superconductivity in YB$_6$. Note, that both the Fermi velocity $v_F$ $\approx$ 1.4$\div$2.1$\times$10$^7$~cm/s and the relaxation time of charge carriers  $\tau_e$ $\approx$ 2.2$\div$2.8$\times$10$^{-14}$~s derived here [Table \hyperref[Tab.3]{III}] are in good agreement with the estimates obtained for YB$_6$ in  $\mu$SR \cite{37} ($v_F$ $\sim$ 10$^7$ cm/s) and in optical conductivity \cite{38} ($\tau_e$ $\approx$ 2.1$\times$10$^{-14}$ s at $T$ = 9~K) studies. Additionally, our temperature variation of $\kappa_1$(0)/$\kappa_1$($T_c$) $\approx$ 1.16 found for all three samples coincides with the result of \cite{3}. At the same time the obtained $\kappa_1$($T$) changes are smaller than these of  $\kappa_2$($T$) and the corresponding ratio is  $\kappa_2$(0)/$\kappa_2$($T_c$) $\approx$ 1.32. The  $\kappa_2$($T$) behavior is practically invariant for all samples changing in contradiction with the previous theoretical and experimental results (see, e.g. \cite{34,35}). Indeed, although according to \cite{34} the differences in the behavior of $\kappa_1$ and  $\kappa_2$ should depend both on the ratio $\xi/l$, and also on the anisotropy of carrier scattering by impurities, but at $T_c$ the equality  $\kappa_1 \approx \kappa_2$ should be valid. Thus, the obtained relation $\kappa_1$($T_c$) $>$  $\kappa_2$($T_c$) in YB$_6$ [Fig.\hyperref[FigX12]{12}] is not consistent with the conclusion of \cite{34}, as well as the changes of the Ginzburg-Landau-Maki parameters  %_________________________TABLE II______
\begin{center}
\begin{longtable}[t]{ccccc}%\firsthline
\caption{Parameters of the superconducting state obtained from magnetization measurements: $T_c$ and  $\Delta$$T_c$ denote the transition temperature and the width of the transition, $H_{cm}$, $H_{c1}$ and $H_{c2}$ the thermodynamic, the first and the second critical fields, $\kappa_2$(0) the Ginzburg-Landau-Maki parameter, $\xi$(0) the coherence length, $\lambda$(0) the penetration depth and $a$ the parameter of relation \hyperref[Eq.10]{(10)}.}\label{Tab.2} \\
\\
\hhline{=====}\\
\quad & \quad\quad & \quad No.1 \quad\quad\quad & \quad No.2 \quad\quad\quad & \quad No.3 \quad\quad\\
\\
\hline\\
\quad $T_c$~(K) & \quad\quad & 7.55  & 7.4  & 6.6  \\
\quad & \quad\quad & & & \\
\quad $\Delta$$T_c$~(K) &\quad\quad & 0.25 & 0.15 & 0.35 \\
\quad & \quad\quad & & & \\
\quad $H_c$~(Oe) & \quad\quad & 615 & 610 & 470 \\
\quad & \quad\quad & & & \\
\quad $H_{c1}$(0)~(Oe) & \quad\quad & 267 & 267 & 147 \\
\quad & \quad\quad & & & \\
\quad $H_{c2}$(0)~(Oe) &\quad\quad & 2902 & 2845 & 3189 \\
\quad & \quad\quad & & & \\
\quad $dH_{c2}$/$dT$~(Oe/K) &\quad\quad & --530 & --530 & --666 \\
\quad & \quad\quad & & & \\
\quad $\kappa_2$(0) &\quad\quad & 3.34 & 3.30 & 4.8 \\
\quad & \quad\quad & & & \\
\quad $\xi$(0)~(${\textmd{\AA}}$) &\quad\quad & 337 & 340 & 321 \\
\quad & \quad\quad & & & \\
\quad $\lambda$(0)~(${\textmd{\AA}}$) &\quad\quad & 1124 & 1121 & 1540 \\
\quad & \quad\quad & & & \\
\quad $a$ &\quad\quad & 0.85 & 0.85 & 0.55 \\%\lasthline
\\
\hhline{=====} \\
\end{longtable}
\end{center}
%_______________________________END TABLE II____________________________

%________________________ TABLE III ______________________
\begin{center}
\begin{longtable}[t]{ccccc}%\firsthline
\caption{Characteristics of charge carriers scattering in YB$_6$: $\rho_0$ -- residual resistivity, $n_e$ -- concentration of charge carriers, $\tau_e$ -- relaxation time of conduction electrons, $v_F$ -- average Fermi velocity, $l$ -- mean free path of conduction electrons.}\label{Tab.3} \\
\\
\hhline{=====}\\
\quad & \quad\quad & \quad No.1 \quad\quad\quad & \quad No.2 \quad\quad\quad & \quad No.3 \quad\quad\\
\\
\hline\\
\quad $\rho_0$~($\mu\Omega$ cm) & \quad\quad & 8.28  & 9.68  & 13.35  \\
\quad & \quad\quad & & & \\
\quad $n_e$ (cm$^{-3}$) &\quad\quad & 1.4$\times10^{22}$ & 1.38$\times10^{22}$ & 1.25$\times10^{22}$ \\
\quad & \quad\quad & & & \\
\quad $\tau_e$ (s) & \quad\quad & 2.8$\times10^{-14}$ & 2.75$\times10^{-14}$ & 2.2$\times10^{-14}$ \\
\quad & \quad\quad & & & \\
\quad $v_F$ (cm/s) & \quad\quad & 2.07$\times10^{7}$ & 2.02$\times10^{7}$ & 1.42$\times10^{7}$ \\
\quad & \quad\quad & & & \\
\quad $l$~(${\textmd{\AA}}$) &\quad\quad & 58 & 55.5 & 31.1 \\%\lasthline
\\
\hhline{=====} \\
\end{longtable}
\end{center}
%________________________ END  TABLE III ______________________

\begin{figure*}[htpb]
\begin{center}
\includegraphics[width = 12.0cm]{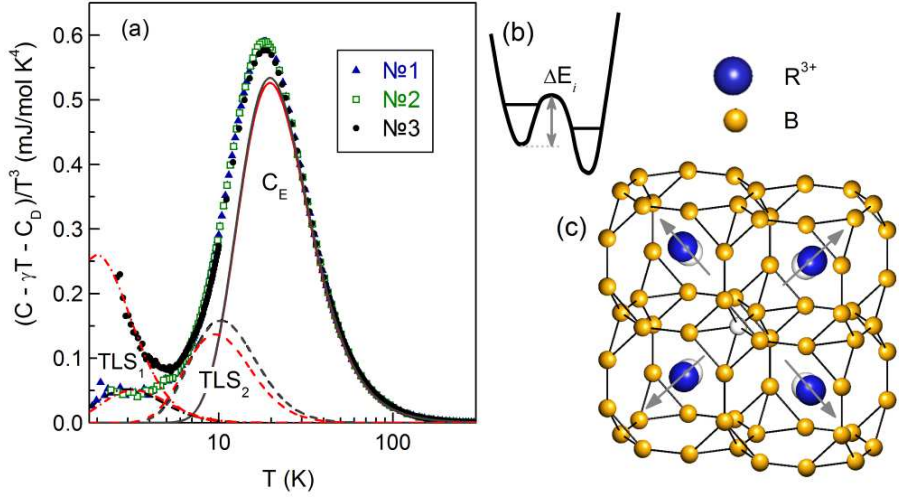}
   \parbox{17cm}{\caption{(Colour on-line) (a) Separation of the contributions to the low temperature heat capacity $(C - \gamma T - C_D)/T^3$ in the normal state ($H_0$ = 5~kOe): the Einstein ($C_E$) component and two types of vacancy (TLS$_1$, TLS$_2$) components are shown by the solid and dashed lines, respectively. (b) Schematic representation of the double-well potential with a barrier height $\Delta E_i$. (c) Crystal structure of YB$_6$. The presence of boron vacancy (shown as a white ball in the center) offsets (represented by arrows) four nearest yttrium $R^{3+}$ ions from their centro-symmetric positions in the cubooctahedrons B$_{24}$.}}\label{FigX14}
   \end{center}
\end{figure*}
\noindent
$\kappa_1$(0)/$\kappa_1$($T_c$) $\approx$ 1.16 and  $\kappa_2$(0)/$\kappa_2$($T_c$) $\approx$ 1.32 are not in accordance with the results \cite{34} of numerical calculations for superconductors in the "dirty limit"
[for YB$_6$ $\xi/l$ = 6$\div$11, see Tables \hyperref[Tab.1]{I}, \hyperref[Tab.2]{II}, \hyperref[Tab.3]{III}]. At the same time, it should be taken into account that the behavior of the thermodynamic $H_{cm}$ and the upper critical field $H_{c2}$ is almost identical for all investigated samples [Fig.\hyperref[FigX13]{13} shows the scaling of $H_{cm}$($T$) and $H_{c2}$($T$) dependences].
Hence, for various YB$_6$ samples distinguished by $T_c$, by residual resistivity and the paramagnetic contribution to magnetic susceptibility [Fig.\hyperref[FigX6]{6}], it is necessary to look for a common mechanism responsible for the decrease of the Ginzburg-Landau-Maki parameter $\kappa_2$ in comparison with the  $\kappa_1$($T$) behavior near $T_c$. The relation between the Ginzburg-Landau-Maki parameters $\kappa_1$($T_c$) $>$  $\kappa_2$($T_c$) found in this study can be explained within the framework of the approach suggested by Maki \cite{28} which additionally takes into account the strong Pauli paramagnetism in the presence of spin-orbit interaction. And, such an unusual relation  $\kappa_1$($T$) $>$  $\kappa_2$($T$) was observed previously by authors of \cite{39} in alloys Nb$_{0.5}$Ta$_{0.5}$ and In$_{0.981}$Bi$_{0.019}$ near $T_c$. Apparently, the emergence and the strengthening in external magnetic field of the spin polarization of electron states can be considered also as the mechanism which is responsible for the different behavior between the $\kappa_1$ and $\kappa_2$ parameters observed in yttrium hexaboride.

At the end of this section, it is worth noting that in accordance with calculations performed in \cite{40} for superconductors with a strong electron-phonon interaction, the parameter $\gamma$$T_c^2$/$H_{cm}^2$(0) depends on the ratio $T_c/\Theta_E$ ($\Theta_E$ --  Einstein temperature) and that this ratio can be used to classify the type of Cooper's pairing. As a result, for YB$_6$ with $T_c/\Theta_E$ $\sim$ 0.075 we found the ratio  $\gamma$$T_c^2$/$\mu_0 V $$H_{cm}^2$(0) $\approx$ 1.7 ($\mu_0$ -- magnetic constant) which is approximately twice lower than the value ($\sim$ 3.7) predicted in \cite{40} for \textit{d}-wave pairing. This supports the expected \textit{s}-type superconductivity in YB$_6$.

\subsection*{\emph{4.2. Characteristics of the normal state of YB$_6$.}}\label{Sec.4p2}
\subsubsection*{\emph{4.2.1 Specific heat at $H =$ 5~kOe.}}

In the analysis of normal state heat capacity contributions of YB$_6$ we used the approach similar to that employed earlier \cite{2},\cite{41}-\cite{45} in studies of higher borides of rare earth elements. In addition to the electronic component $C_{el}$ = $\gamma T$ with $\gamma$ $\approx$ 3.8~mJ/(mol K$^2$) and the Debye $C_D$ contribution which originates from the rigid covalent framework of boron atoms
\begin{equation}\label{Eq.13}
C_D = 9 r R \left(\frac{T}{\Theta_D}\right)^3 \int\limits_0^{\Theta_D/T} e^x x^4 \left[ e^x - 1\right]^{-2} dx
\end{equation}
(for $R$B$_6$ $r$ = 6, R is the universal gas constant, $\Theta_D$ the Debye temperature), in this case also the Einstein $C_E$ component of the specific heat
\begin{equation}\label{Eq.14}
C_E = 3 R N_E \left(\frac{\Theta_E}{T}\right)^2 exp\left(\frac{\Theta_E}{T}\right) \left[exp\left(\frac{\Theta_E}{T}\right) - 1 \right]^{-2}
\end{equation}
($N_E$ -- number of oscillators per unit cell), has to be taken into account. The $C_E$ term \hyperref[Eq.14]{(14)} is caused by quasi-local vibrations of yttrium ions located in the cavities 

%______________  TABLE IV  _____________
\begin{longtable}[t]{ccccc}%\firsthline
\caption{Parameters of the heat capacity $(C - \gamma T - C_D)$/$T^3$ = \textit{f}($T$) [Fig.\hyperref[FigX14]{14}] approximation by formulas \hyperref[Eq.14]{(14)} and \hyperref[Eq.15]{(15)}: $\Delta E_1$,  $\Delta E_2$ and $N_1$, $N_2$ are the height of barriers and the reduced concentrations of double-well potentials TLS$_1$ and TLS$_2$, respectively, $N_E$ is the reduced concentration of Einstein oscillators. We present also the chemical composition found for the investigated YB$_6$ crystals, and the mass densities $g_m^C$ and $g_m$ obtained from $C$($T$) and hydrostatic measurements, respectively.}\label{Tab.4} \\
\\
\hhline{=====}\\
\quad & \quad\quad &  No.1 \quad\quad & No.2 \quad\quad &  No.3 \quad\\
\\
\hline\\
\quad $\Delta$$E_2$ (K) & \quad\quad & 51.2 \quad\quad & 51.4 \quad\quad & 47.9 \quad \\
\quad & \quad\quad & &\quad\quad & \quad\\
\quad $N_2$ = 4$n_v$ & \quad\quad & 0.123 \quad\quad & 0.126 \quad\quad & 0.106 \quad \\
\quad & \quad\quad & &\quad\quad & \quad\\
\quad $\Delta$$E_1$ (K) & \quad\quad & 13.9 \quad\quad & 13.3 \quad\quad & 10.8 \quad \\
\quad & \quad\quad & &\quad\quad & \quad\\
\quad $N_1$ $\sim$ $n_d$ & \quad\quad & 0.00073 \quad\quad & 0.00076 \quad\quad & 0.00357 \quad \\
\quad & \quad\quad & &\quad\quad & \quad\\
\quad $N_E $ & \quad\quad & 0.95 \quad\quad & 0.945 \quad\quad & 0.96 \quad \\
\quad & \quad\quad & &\quad\quad & \quad\\
\quad Chemical & \quad\quad & &\quad\quad & \quad\\
\quad composition & \quad\quad & Y$_{0.95}$B$_{5.816}$ \quad\quad & Y$_{0.945}$B$_{5.811}$ \quad\quad & Y$_{0.96}$B$_{5.841}$ \quad\\
\quad  from $C$($T$)& \quad\quad & &\quad\quad & \quad\\
\quad & \quad\quad & &\quad\quad & \quad\\
\quad $g_m^C$ (g/cm$^3$) & \quad\quad & 3.55 \quad\quad & 3.54 \quad\quad & 3.58 \quad \\
\quad & \quad\quad & &\quad\quad & \\
\quad $g_m$ (g/cm$^3$)  & \quad\quad & 3.559$\pm$0.006 \quad\quad & 3.56$\pm$0.01 \quad\quad & 3.62$\pm$0.02 \quad \\%\lasthline
\\
\hhline{=====} \\
\end{longtable}
%______________  TABLE IV  _____________
\noindent
formed by boron B$_{24}$ 
cubooctahedra and loosely bound to the rigid covalent boron sub-lattice. Eqs.\hyperref[Eq.13]{(13)} and \hyperref[Eq.14]{(14)} allowed us to estimate Einstein ($\Theta_E$ $\approx$ 97.2~K) and Debye ($\Theta_D$ $\approx$ 1160~K) temperatures for YB$_6$. For example, Fig.\hyperref[FigX14]{14(a)} presents in coordinates $(C - \gamma T - C_D)/T^3$  vs. $T$ the low temperature heat capacity of investigated YB$_6$ crystals. The received value $\Theta_E$ $\approx$ 97.2~K is consistent with the results of point-contact \cite{29} and tunneling \cite{30} spectroscopy measurements ($\omega_E$ $\sim$ 8~meV), and with inelastic neutron scattering ($\omega_E$ $\sim$ 10~meV) \cite{46} and Raman scattering ($\omega_E$ $\sim$ 70~cm$^{-1}$) \cite{47} data. The value $\Theta_D$ $\approx$ 1160~K coincides with the results of heat capacity calculations for LaB$_6$ \cite{41}-\cite{44}. It is also in agreement with the data of $\Theta_D$ $\approx$ 1160--1190~K obtained in \cite{2,45,48} for the analog non-magnetic higher boride $-$ lutetium dodecaboride (LuB$_{12}$), and comparable to $\Theta_D$ $\approx$ 1250--1370~K deduced for $\beta$-boron in X-ray diffraction studies \cite{48}.

Along with Einstein component, which leads to maximum on $(C - \gamma T - C_D)/T^3$ vs. $T$ curves near 20~K [see Fig.\hyperref[FigX14]{14(a)}], we observed two additional features on these dependence $-$ one near 10~K and another below 4~K. The separation of these low-temperature contributions was made in the same manner as it was
done for LaB$_6$ in \cite{43,44}, where the heat capacity below 20~K was associated with two additive two-level components attributed to vibrations of rare earth ions in the vicinity of boron vacancies (see two-level systems TLS$_1$ and TLS$_2$ in Fig.\hyperref[FigX14]{14(a)}, and also \cite{49}). These two TLS terms were described by the Schottky formula
\begin{eqnarray}\label{Eq.15}
C_{Sh_i} = R N_i g_{0i} g_{1i} \left(\frac{\Delta E_i}{T}\right)^2 exp\left(\frac{\Delta E_i}{T}\right)\nonumber\\
 \times \left[ g_{0i} exp\left(\frac{\Delta E_i}{T}\right) + g_{1i} \right]^{-2},
\end{eqnarray}
where $g_{0i}$, $g_{1i}$ denote the degeneracy of the ground and excited states, $\Delta E_i$ the splitting energy and $N_i$ the concentration of the two-level systems (TLS). The analysis of the low temperature heat capacity $(C - \gamma T - C_D)/T^3$ of LaB$_6$ was undertaken in \cite{43,44} in the framework of relation \hyperref[Eq.15]{(15)} for three different schemes of levels, including \textbf{3--1}, \textbf{1--1} and \textbf{1--3} configurations. As a result, describing the heat capacity of La$^\textmd{N}$B$_6$ with various boron isotopes ($N$ = 10, 11, \textit{nat}) authors \cite{43,44} choose TLS diagrams consisting of singlet and triplet states, which allowed them to obtain the best fit with the lowest concentration of boron vacancies in the hexaboride compounds.

We emphasize that the presence of boron vacancies has been clearly confirmed in X-ray and neutron studies of $R$B$_6$, and it was shown in \cite{50}-\cite{53} that there are about 1--9$\%$ of vacancies in the boron sub-lattice in all known hexaborides. The concentration of these defects depends both on the $R$--ion and the method of the single crystal growth. The presence of boron vacancies, on one side, and the loosely bound states of $R$-ions in the rigid covalent boron framework, on the other side, lead to displacements of the $R^{3+}$--ions from their centro-symmetric positions inside the truncated B$_{24}$ cubooctahedra [see Fig.\hyperref[FigX14]{14(c)}]. This gives rise to a disorder in the arrangement of yttrium ions in the hexaboride matrix. The derangement increases with temperature lowering, and depending on the concentration of intrinsic defects and impurities, a number of non-equivalent positions is expected for $R^{3+}$ ions in $R$B$_6$. Thus, similar to amorphous compounds and glasses \cite{54}, in the cage-glass configuration the appearance of two level systems is related to the disorder in the arrangement of heavy ions in $R$B$_6$ crystals. In other words, the emergence of TLS seems to be equivalent to the formation of different double-well potentials with a barriers $\Delta E_i$ [Fig.\hyperref[FigX14]{14(b)}].

Following the approach, the low-temperature data $C/T^3$ of YB$_6$ [Fig.\hyperref[FigX14]{14(a)}] have been approximated by Eqs.\hyperref[Eq.14]{(14)} and \hyperref[Eq.15]{(15)} with an Einstein contribution $C_E$ ($\Theta_E$ $\approx$ 97.2~K) and two types of two-level systems TLS$_i$ ($i$ = 1, 2) each consisting of singlet ($g_{0i}$ = 1) and triplet ($g_{1i}$ = 3) states. The barrier height of TLS$_2$ [see Fig.\hyperref[FigX14]{14(b)} and Table \hyperref[Tab.4]{IV}] was found to be $\Delta E_2$ $\sim$ 50~K, and this value does not practically depend on the concentration of intrinsic defects in crystals No.1--No.3. The obtained relative concentration  of cells with a double-well

%_______________ TABLE  V__________________
\LTcapwidth=16cm
\begin{longtable*}[t]{ccc}
\caption{Parameters of the magnetic susceptibility  $\chi$(T) approximation by Eq.\hyperref[Eq.16]{(16)}: $N_{m0}$  -- concentration of magnetic centers per unit cell, $\mu_{\textmd{eff}}$ -- the effective magnetic moment of magnetic centers, $\chi_d$ -- diamagnetic contribution to susceptibility. The concentration of ytterbium impurities $x$(Yb) in YB$_6$ samples detected from spectral analysis data are also presented. }\label{Tab.5} \\
\hhline{===} \\
\quad\quad\quad\quad  & & $\quad\chi$($T$, $H_0$ = 5.4~kOe) \quad\quad\quad\quad \\
%\\
%\\
%\multicolumn{3}{c}{}\\
\begin{tabular}[t]{ccc}%\firsthline
\hline \\
\quad\quad YB$_6$&\quad\quad & $x_{\textmd{Yb}}$ (ppm)\quad \\
 \\
\hline\\
%\multicolumn{3}{c}{}\\
%\multicolumn{3}{c}{}\\
\quad\quad No.1 &\quad & 10 \quad\\
\quad\quad No.2 & \quad& 10  \quad\\
\quad\quad No.3 &\quad & 1000 \quad\\%\lasthline
\end{tabular} & \quad\begin{tabular}[t]{c}%\firsthline
%\hline\\
\\
\\
%\hline\\
\\
\\
 \\
 \\
 \\
\end{tabular} & \begin{tabular}[t]{ccccccc}%\firsthline
\hline \\
\quad$N_{m0}$ $\mu_{\textmd{eff}}^2$  (emu/mol)& \quad\quad & $N_1$ = $N_{m0}$ (f.u.) & \quad\quad & $\mu_{\textmd{eff}}$ ($\mu_{B}$) &\quad\quad & $\chi_d$$\times$10$^5$ (emu/mol)\quad\quad \\
 \\
\hline\\
%\multicolumn{3}{c}{}\\
%\multicolumn{3}{c}{}\\
\quad 0.00344 & \quad\quad & 0.00073 & \quad\quad & 2.17 & \quad\quad & -- 3.8  \quad\quad \\
\quad 0.00288 & \quad\quad & 0.00076  & \quad\quad & 1.95 & \quad\quad  & -- 4.8  \quad\quad \\
\quad 0.0144    & \quad\quad & 0.00357    & \quad\quad & 2.01  & \quad\quad  & -- 2.6 \quad\quad  \\
\end{tabular} \quad\\
\\
\hhline{===}
\end{longtable*}
%_______________ TABLE  V__________________

\noindent
potential was found to be $N_2$ $\approx$ 0.106$\div$0.126, and similar as in the case of LaB$_6$ \cite{43,44}, it should be associated with the number of boron vacancies in the YB$_6$ structure. Each vacancy namely produces a displacement of the yttrium ions from their centro-symmetric positions in four neighboring B$_{24}$ clusters [see Fig.\hyperref[FigX14]{14(c)}], so the genuine concentration of boron vacancies should be $n_v$(B) = $N_2$/4 $\approx$ 2.6$\div$3.2 $\%$, indicating also a notable decrease of mass density in studied YB$_6$ single crystals. On the other side, the observed concentration of TLS$_1$ in crystals No.1 and No.2 [$\Delta E_1$ $\approx$ 10$\div$14~K, $N_1$ $\approx$ 0.73$\div$0.76 $\times$10$^{-3}$, Table \hyperref[Tab.4]{IV}] is quite small when compared with the amount of boron di-vacancies previously detected in LaB$_6$ \cite{43,44}. Indeed, the estimation of the concentration of boron di-vacancies in case of their random distribution in the $R$B$_6$ matrix leads to $n_d$(B) = $n_v$(B)(1 $-$ [1 $-$ $n_v$(B)]$^z$) $\approx$ 3.5 $\times$10$^{-3}$ (where $z$ = 4 is the coordination number in the boron sub-lattice), which with a good accuracy corresponds to the TLS$_1$ concentration $N_1$ found for sample No.3 [Table \hyperref[Tab.4]{IV}]. On the contrary, low values of TLS$_1$ concentrations $N_1 \ll n_d$(B) for crystals No.1 and No.2, seem to indicate a decrease of the number of di-vacancies in favor of single vacancies [see Table \hyperref[Tab.4]{IV}], that evidenced opposite their random distribution in YB$_6$.
Note that a number of yttrium vacancies [n$_v$(Y) = 1 $-$ $N_E$ $\approx$ 4$\div$5.5~$\%$, Table \hyperref[Tab.4]{IV}] is also detected from the heat capacity analysis and it contributes to the mass density lowering in YB$_6$.

\subsubsection*{\emph{4.2.2 Analysis of magnetic susceptibility.}}

The analysis of contributions to magnetic susceptibility of samples No.1--No.3 in the normal state [Fig.\hyperref[FigX6]{6}] was carried out in the framework of relation
\begin{equation}\label{Eq.16}
\chi = M/H = N_{m0} \mu_{\textmd{eff}}^2/\left(3 k_B T \right) + \chi_d,
\end{equation}
where $N_{m0}$ is the concentration of magnetic centers in small magnetic fields, $\Theta_p$ the paramagnetic Curie temperature and $\chi_d$ the diamagnetic susceptibility. Fig.\hyperref[FigX6]{6} shows the fitting results of the experimental curves  $\chi$($T$) by Eq.\hyperref[Eq.16]{(16)} indicating that within the limits of experimental accuracy the susceptibility follows the Curie-Weiss dependence. Table \hyperref[Tab.5]{V} presents the parameters obtained by this approximation.

It is worth noting that the localized magnetic moments with concentrations of $N_{m0}$, determine the paramagnetic susceptibility of YB$_6$ (Fig.\hyperref[FigX6]{6}), correspond to small fields [$H$ = 5.4~kOe, linear $M$($H$) dependence]. It was found from the optical emission spectral analysis that the magnetic impurity doping level in the samples No.1--No.2 is about 10~ppm. Hence, in the absence of magnetic impurities the detected magnetic moments may be associated with complexes of vacancies in the matrix of this nonmagnetic hexaboride. It should be mentioned that although in strong magnetic fields the low temperature magnetic contribution to heat capacity within the experimental accuracy could not be clearly detected in crystals No.1 and No.2 [Fig.\hyperref[FigX5]{5(a)}], the presence of a small amount of complexes of vacancies ($N_1$ = 730$\div$760~f.u.$\sim$ 100~ppm, see Tables \hyperref[Tab.4]{IV} and \hyperref[Tab.5]{V}) in these two samples can account for their Curie-Weiss dependence of magnetic susceptibility $\chi$($T$) [Fig.\hyperref[FigX6]{6}]. In this case the strong increase ($\sim$ 5 times) of concentration of divacancies $N_1$ in sample No.3 ($N_1$ = 3570~f.u.$\sim$500~ppm, Table \hyperref[Tab.4]{IV}) compared with No.1 and No.2 leads in small fields to a proportional elevation of the paramagnetic response [Fig.\hyperref[FigX6]{6} and Table \hyperref[Tab.5]{V}]. Thus, within the approach the complexes of vacancies in the YB$_6$ matrix are responsible both for the appearance of the low temperature heat capacity component [TLS$_1$ in Fig.\hyperref[FigX14]{14(a)}] and for the emergence of the paramagnetic Curie-Weiss term in the magnetic susceptibility $\chi$($T$, $H_0$ = 5.4~kOe) [Fig.\hyperref[FigX6]{6}]. Taking the concentration $N_{m0}$ = $N_1$ from the analysis of heat capacity at $H$ = 5~kOe and the $N_{m0}$$\mu_{\textmd{eff}}^2$ parameter obtained from the susceptibility approximation by relation \hyperref[Eq.16]{(16)}, we can estimate the value of the magnetic moment of these magnetic complexes  $\mu_{\textmd{eff}}$ = 1.95$\div$2.17~$\mu_B$ for No.1--No.3 crystals [Table \hyperref[Tab.5]{V}].

The calculated value of the magnetic moment $\mu_{\textmd{eff}}$ $\approx$ 2.36~$\mu_B$ obtained in \cite{55} for  clusters of boron vacancies in $R$B$_6$ serve in favor of this interpretation. It is worth noting also that in \cite{56} weak magnetic states were predicted in two-dimensional boron composed of B$_{20}$ clusters in a hexagonal arrangement. In addition, suppression of superconductivity due to the formation of magnetic moments in the vicinity of nonmagnetic Lu impurities has been recently found in Zr$_{1-x}$Lu$_x$B$_{12}$ \cite{57} and associated with the spin polarization of \textit{d}-states in the conduction band. In favor of this alternative spin-polaron scenario points e.g. the weak ferromagnetism of charge carriers observed both in some nonmagnetic hexaborides as Ca$_{1-x}$La$_x$B$_6$, Ca$_{1-x}$Ba$_x$B$_6$ \cite{58,59} and in paramagnetic phase of PrB$_6$ \cite{60,61}.

It should be mentioned here that contrary to the case of samples No.1, No.2, the spectral analysis shows that the crystal No.3 contains 200--1000~ppm of ytterbium impurities. Thus, compared with samples No.1, No.2, a large magnetic contribution to the heat capacity and magnetic susceptibility of crystal No.3 may at first glance be associated with magnetic Yb$^{3+}$ impurities. However, it was shown in \cite{62}-\cite{64} that Yb-ions are divalent and nonmagnetic in the $R$B$_6$ matrix. Moreover, in diamagnetic YbB$_6$ compound the concentration of magnetic Yb$^{3+}$ ions is very small and varies within 0.1$\div$2~$\%$. Thus, even if taking into account the presence of about 1000~ppm of ytterbium impurities (the upper limit found by the spectral analysis in sample No.3), there seem to be no more than 20~ppm of magnetic Yb$^{3+}$ centers. As a result, the estimated concentration of magnetic centers 730$\div$3570~f.u. (= 104$\div$510~ppm) with effective moment $\mu_{\textmd{eff}}$ $\approx$ 2~$\mu_B$ serves as an argument against the explanation of the magnetic contribution to heat capacity and of magnetization in terms of YB$_6$ doping by Yb$^{3+}$ magnetic impurities. Moreover, it also seems to allow excluding a direct correlation between the concentration of ytterbium impurities and the value elevation of residual resistivity in YB$_6$.  At the end of this section it is worth noting that in our experimental study also attempts were undertaken to measure the field dependence of magnetization of all No.1--No.3 crystals with the help of the PPMS-9. However, the signal from sample holder which was comparable with the magnetization of sample No.3 did not allowed us to carry out the separation and analysis of contributions in strong magnetic fields.

\subsubsection*{\emph{4.2.3 Anomalies of charge transport and of thermodynamic parameters near $T^*$.}}

The formation of two-level systems in higher borides was for the first time observed experimentally in rare earth dodecaborides LuB$_{12}$ \cite{2} and ZrB$_{12}$ \cite{45} which are composed of a rigid framework formed by boron B$_{12}$ nanoclusters and heavy ions embedded in cavities arranged by B$_{24}$ cubooctahedra. In Raman spectra \cite{2} of single crystalline Lu$^\textmd{N}$B$_{12}$ samples with a different isotopic composition of boron ($N$ = 10, 11, \textit{nat}) it was shown that the Raman response exhibits a boson peak at liquid nitrogen temperatures and such a feature  in the low-frequency range is a fingerprint of systems with strong structural disorder. To explain the properties of LuB$_{12}$ authors of \cite{2} have proposed a model of cage-glass formation with a phase transition at $T^*$ $\sim$ 50--70~K, and it was found that the barrier height of the double-well potential $\Delta E$ [Fig.\hyperref[FigX14]{14(b)}] is practically equal to the cage-glass transition temperature $T^*$. At the same time the temperature lowering at $T < T^*$ leads to displacements of metallic ions from their centro-symmetric positions inside the B$_{24}$ cubooctahedra [see e.g. Fig.\hyperref[FigX14]{14(c)} for $R$B$_6$]. The result is a static disorder in the arrangement of $R^{3+}$ ions while maintaining the rigid covalent boron framework. The presence of two-level systems with a barrier $\Delta E$ $\sim$ 90~K was reliably demonstrated also in LaB$_6$ and in Ce$_x$La$_{1-x}$B$_6$ solid solutions based on low-temperature heat capacity measurements \cite{43,44}. Furthermore, a pseudo-gap \cite{65} and a low-frequency peak in inelastic light scattering spectra \cite{66,67} were found in LaB$_6$. Taking into account that the ionic radius of yttrium $r_i$(Y$^{3+}$) $\sim$ 0.92~${\textmd{\AA}}$ is significantly lower than that of La$^{3+}$ ($r_i$ $\sim$ 1.17~${\textmd{\AA}}$), which points to loosely bound states of Y$^{3+}$ ions in the boron sub-lattice, a strong non-equilibrium state with a considerable structural disorder together with formation of TLSs having a low barrier height can be expected for YB$_6$. Moreover, the observed ratio B/Y (about 6.1 \cite{9}) in the YB$_6$, which is large compared to the stoichiometric value for hexaborides, suggest the presence of a large number of vacancies in the sub-lattice of yttrium which prevails the vacancy concentration on boron sites.

The above estimated barrier height of the double-well potential TLS$_2$ $\Delta E_2$ $\approx$ 50~K [Table \hyperref[Tab.4]{IV}] should be therefore related to the glass transition temperature $T^*$ $\sim$ 50~K which corresponds to the occurrence of structural disorder in the subsystem of Y$^{3+}$ ions in YB$_6$. At the same time also features found in the vicinity of $T^*$ $\sim$ 50~K on resistivity derivatives [Fig.\hyperref[FigX2]{2(c)}], Hall coefficient $R_H$($T$) [Fig.\hyperref[FigX9]{9}] and on Seebeck coefficient $S$($T$) [Fig.\hyperref[FigX10]{10}] should be considered as anomalies that arise near this phase transition. It is worth noting that unlike to cage-glasses LuB$_{12}$ and LaB$_6$ in which high charge carrier motilities $\mu_H$ $\sim$ 2500~cm$^2$/(V s) \cite{68} and $\sim$ 21000~cm$^2$/(V s) \cite{69} were observed, respectively, in YB$_6$ the proximity to lattice instability and the resultant stronger structural disorder cause a dramatic suppression of Hall mobility. In the samples No.1, No.2 and No.3 with different concentrations of boron vacancies $n_v$ and of paramagnetic centers $N_1$ = $N_{m0}$ [see Tables \hyperref[Tab.4]{IV} and \hyperref[Tab.5]{V}] the low-temperature mobility values are very similar and they do not exceed 50~cm$^2$/(V s) [insets in Fig.\hyperref[FigX9]{9}]. This mobility value corresponds to a very small value of the relaxation time $\tau_e$ = 2.2$\div$2.8$\times$10$^{-14}$~s and of the mean free path of the charge carriers $l$ = 31$\div$58~${\textmd{\AA}}$ [Table \hyperref[Tab.3]{III}]. Additionally, the close to square root power-law dependence $\mu_H$ $\sim$ $T^{-\alpha}$ ($\alpha$ $\sim$ 0.5) of the mobility in YB$_6$ [see insets in Fig.\hyperref[FigX9]{9}] is significantly weaker than those observed in LuB$_{12}$ ($\alpha$ $\sim$ 2.06) \cite{68} and LaB$_6$ ($\alpha$ $\sim$ 3) \cite{69}. It is also worth noting that substantial structural distortions in YB$_6$ were detected by low-temperature Raman studies \cite{47}.

\begin{figure}[t]
\begin{center}
\includegraphics[width = 7.4cm]{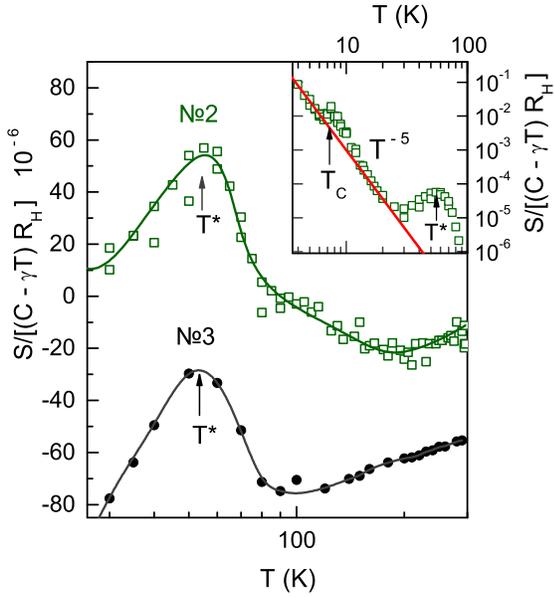}
    \parbox{8cm}{\caption{(Colour on-line) Temperature dependences of the parameter $S/\left[(C - \gamma T) R_H \right]$ for samples No.2 and No.3 of YB$_6$. The inset shows the approximation of data by exponential dependence $\sim T^{-\gamma}$ (see text). Arrows indicate the superconducting transition ($T_c$) and the cage-glass transition ($T^*$) temperatures.}}\label{FigX15}
   \end{center}
\end{figure}

To analyze the features of charge transport and of thermodynamic characteristics near $T^*$ the relation
\begin{equation}\label{Eq.17}
S_{ph} = (C_{ph}/n e)\left[1 + \tau_{e-ph}/\tau\right]^{-1},
\end{equation}
which connects the Hall coefficient $R_H = 1/ne$, the phonon drag thermopower $S_{ph}$ and the phonon contribution to the heat capacity $C_{ph}$ \cite{70}, was used [in relation \hyperref[Eq.17]{(17)} $\tau_{e-ph}$ and $\tau$  denote the electron-phonon relaxation time and the relaxation time of the phonon gas, respectively]. Using the experimental results of Figs.\hyperref[FigX3]{3}, \hyperref[FigX9]{9} and \hyperref[FigX10]{10} it is then possible to obtain an estimation of the temperature dependence of the factor $\left[1 + \tau_{e-ph}/\tau \right]^{-1}$ = $S/\left[(C - \gamma T) R_H \right]$, which determines the relative change of relaxation times in the system of conduction electrons. It can be seen in Fig.\hyperref[FigX15]{15} that the parameter $\left[1 + \tau_{e-ph}/\tau \right]^{-1}$ of YB$_6$ samples passes through a maximum in the vicinity of $T^*$ $\sim$ 50~K and that the cage-glass transition temperature $T^*$ corresponds to a sharp change in charge carriers scattering. In the cage-glass phase at temperatures below 20 K, where charge scattering by impurities and structural defects becomes dominant [$\rho_0$ $\approx$ \textit{const} and $\tau_{e-ph}$ $\approx$ \textit{const}, see Fig.\hyperref[FigX2]{2(a)} and insets in Fig.\hyperref[FigX9]{9}], we observed a strong power-law dependence of $\left[1 + \tau_{e-ph}/\tau \right]^{-1}$ $\sim$ $T^{-\gamma}$ with $\gamma$ $\sim$ 5 which should correspond to the scattering in the phonon subsystem \cite{71}. A detailed quantitative analysis of the charge transport anomalies requires a correct separation of two contributions to Seebeck coefficient $-$ the negative (Mott-type) diffusion thermopower and the phonon drag effect $-$ which is beyond the scope of this paper and will be published elsewhere. At the same time, the temperature dependence of $\left[1 + \tau_{e-ph}/\tau \right]^{-1}$ = $S/\left[(C - \gamma T) R_H \right]$ (Fig.\hyperref[FigX15]{15}) allows to confirm that in YB$_6$ there are two phase transitions $-$ at $T^*$ $\sim$ 50~K and at $T_c$ $\sim$ 6$\div$7.6~K $-$ into the cage-glass state and into the superconducting state, correspondingly.

\subsubsection*{\emph{4.2.4 Factors responsible for the $T_c$ dispersion.}}

To quantify the offset from the equilibrium state in YB$_6$ one can use parameters $N_E$ and $N_2$ which were found from the heat capacity analysis shown in Fig.\hyperref[FigX14]{14} by Eq.\hyperref[Eq.14]{(14)} and \hyperref[Eq.15]{(15)}, and determine the number of vacancies in Y and B sub-lattices, respectively. The concentration of vacancies of yttrium $n_v$(Y) = 1 $-$ $N_E$ = 4$\div$5.5~$\%$ and boron $n_v$(B) = $N_2$/4 $\approx$ 2.6$\div$3.2~$\%$ of samples under investigation are given in Table \hyperref[Tab.4]{IV}. The resulting chemical composition [see Table \hyperref[Tab.4]{IV}] is then $\sim$ Y$_{0.95}$B$_{5.81}$ [for samples No.1 and No.2] and Y$_{0.96}$B$_{5.84}$ [for sample No.3], and provide a Y/B ratio in the range of 6.08$\div$6.14. This ratio is in good agreement with the abovementioned results of \cite{9} ($\sim$ 6.1) and allows to link the variations of $T_c$ with significant deviations from the hexaboride stoichiometry both in the yttrium and boron subsystem. Thus, taking into account the X-ray density $g_m$ = 3,705 g/cm$^3$, which within experimental accuracy remains equal for all investigated YB$_6$ crystals, one can estimate from the heat capacity analysis made above [see Fig.\hyperref[FigX14]{14}] the mass density values $g_m^C$ = 3.54$\div$3.58 g/cm$^3$ of samples No.1--¹3 [see Table \hyperref[Tab.4]{IV}] by the relation
\begin{equation}\label{Eq.18}
g_m^C = \frac{([1 - n_v(\textmd{Y})] m_\textmd{Y} + 6 [1 - n_v(\textmd{B})]m_\textmd{B}) 1.66057\times10^{-24}}{a^3\times10^{-24}},
\end{equation}
where $a$ is the lattice parameter, and $m_\textmd{Y}$ and $m_\textmd{B}$ denote the atomic mass of yttrium and boron, correspondingly. Results of independent hydrostatic density measurements of No.1--No.3 crystals are also shown in Table \hyperref[Tab.4]{IV}. As can be seen from Table \hyperref[Tab.4]{IV}, the parameters $g_m^C$ and $g_m$ are in good agreement with each other, and, as expected, the lower density observed for samples No.1, No.2 meets the higher concentration of single vacancies on yttrium and boron sites.

From these results a direct correlation between the development of the structural instability in YB$_6$ crystals and their transition temperature $T_c$ can be seen. Namely, the non-equilibrium state is developed in No.1 and No.2 with a higher concentration of sole vacancies in yttrium and boron lattices, corresponding to higher transition temperature values $T_c$ = 7.4$\div$7.55~K. On the other side, the lower transition temperature value $T_c$ = 6.6~K observed in crystal No.3 corresponds to a lower concentration of Y and B vacancies. Moreover, in this sample there is an additional superconductivity suppression mechanism associated with Cooper pairs breaking by scattering on localized magnetic moments of vacancy complexes. In this case, when comparing the superconducting characteristics of YB$_6$ samples [Tables \hyperref[Tab.1]{I}, \hyperref[Tab.2]{II}], we can note that crystal No.3 with the highest concentration of induced magnetic moments has the smallest $T_c$ and $H_{cm}$ values, but the upper critical field $H_{c2}$(0) and the derivative $d H_{c2}/dT$($T_c$) of this sample are the highest. Sample No.3 exhibits also the highest values of Ginzburg-Landau parameters $\kappa_1$ and $\kappa_2$ [Fig.\hyperref[FigX12]{12}], together with the lowest amplitude of the $\Delta C$ jump near $T_c$ [Table \hyperref[Tab.1]{I}] and a significant broadening of the heat capacity anomaly [Fig.\hyperref[FigX4]{4}].

Considering a significant softening of the low-frequency phonon modes with temperature lowering in YB$_6$ found in \cite{72}, we may expect a relation between the softening and the development of lattice instability, leading to a $T_c$ increase. Authors of \cite{72} pointed out that just above the transition to superconducting state the low-frequency branches in Raman spectra exhibit energies of $\sim$ 42~cm$^{-1}$ ($\sim$ 5~meV) and $\sim$ 60$\div$70~cm$^{-1}$ ($\sim$ 8~meV). Similarly, when creating the Eliashberg function $\alpha^2$($\omega$)$F$($\omega$) from tunneling spectra measured on YB$_6$ single crystals with $T_c$ $\approx$ 7.1~K, two features were found on $\alpha^2$($\omega$)$F$($\omega$) in \cite{30} $-$ a "shoulder" at 4.9~meV and a broad peak in the vicinity of 8.5~meV. In addition, authors of \cite{30} estimated the electron-phonon interaction constant $\lambda_{e-ph}$ $\approx$ 0.9. Similar values of the energy of Einstein oscillators $\Theta_{E1}$ $\approx$ 51~K ($\sim$ 4.5~meV) and  $\Theta_{E2}$ $\approx$ 90~K ($\sim$ 8~meV) in YB$_6$ were obtained from the analysis of phonon heat capacity in \cite{3}. Taking into account the results of the present study, it seems to be reasonable that the above features observed experimentally at $\sim$ 50~K should be associated with the barrier value of the double-well potential $\Delta E_2$ [Table \hyperref[Tab.4]{IV}] which corresponds to the cage-glass transition temperature $T^*$ $\approx$ $\Delta E_2$/$k_B$ $\approx$ 50~K. Thus, in accordance with the conclusions of \cite{3} and \cite{30}, the formation of Cooper pairs in YB$_6$ occurs through the electron-phonon interaction with quasi-local vibrations of yttrium ions with energies $\Theta_E$ $\approx$ 8~meV. Using parameters  $\omega_{\textmd{ln}}$ $\approx$ $\Theta_E$ $\approx$ 97.2~K, $\lambda_{e-ph}$ $\approx$ 0.96 obtained in this work in the strong coupling limit and taking the Coulomb pseudopotential $\mu^*$ $\sim$ 0.07, within the framework of the relation for superconducting transition temperature \cite{73}
\begin{equation}\label{Eq.19}
k_B T_c = \frac{\hbar\omega_{\textmd{ln}}}{1.2} exp\left[- \frac{1.04 (1 + \lambda_{e-ph})}{\lambda_{e-ph} - \mu^* (1 + 0.62 \lambda_{e-ph})} \right]
\end{equation}
we obtain $T_c$ $\approx$ 7.3~K, which correlates very well with $T_c$ = 7.4$\div$7.55~K observed for crystals No.1 and No.2 of YB$_6$.

\subsubsection*{\emph{4.2.5 Residual resistivity in YB$_6$.}}

To explain the lower values of residual resistivity $\rho_0$ in samples No.1, No.2 where higher $T_c$ values were detected [Fig.\hyperref[FigX2]{2}] in combination with the higher concentration of boron and yttrium vacancies [Table \hyperref[Tab.4]{IV}], a mechanism associated with vacancy ordering in the matrix of yttrium hexaboride may be considered. Ordered structures of vacancies have been already observed e.g. in the family of MNiSn (M - Ti, Zr, Hf) compounds \cite{74,75} and it was shown that the appearance of vacancy superstructures leads in some cases to extremely low values of resistivity of these compounds. Under such an approach, a change in the random distribution of boron vacancies and the formation of structures of single vacancies in the YB$_6$ matrix [in samples No.1 and No.2] can be associated with a decrease of electron scattering. On the other hand, scattering on impurities and vacancy complexes [in sample No.3] leads to a higher residual resistivity. Vacancy complexes can be namely considered as non-point defects which provide significant structural distortions and are also related to magnetic moments of $\sim$ 2~$\mu_B$ [see Table \hyperref[Tab.5]{V}]. Structural distortions near these clusters of defects, along with a $\rho_0$ increase, lead also to an increase of heat capacity and to broadening of $C$($T$) features' at $T_c$ seen in crystal No.3 [Fig.\hyperref[FigX4]{4}]. We emphasize that in sample No.3 we observed the highest concentration of two-level-systems (TLS$_1$) $N_1$, which $\sim$ 5 times exceeds the $N_1$ values in crystals No.1 and No.2 [Table \hyperref[Tab.4]{IV}] as well as a paramagnetic signal which is $\sim$ 5 times higher than the $\chi$($T$) values of No.1 and No.2. It was mentioned in \cite{3} that high temperature annealing of YB$_6$ samples always leads to a suppression of superconductivity. Within our approach this can be associated with the destruction of the lattice of sole vacancies, with the forming of clusters of these defects and the emergence of new paramagnetic centers on their basis.

\section*{V. CONCLUSIONS}\label{Sec.5}

A correlation of experimental results obtained from resistivity, Hall and Seebeck coefficients, heat capacity and magnetization measurements of YB$_6$ allowed us to observe for the first time the transition into an unusual cage-glass state at $T^*$ $\sim$ 50~K in which the yttrium ions are displaced from their central positions in boron B$_{24}$ cubooctahedra and located randomly in these cavities of the rigid covalent boron sub-lattice. We have shown that the number of isolated vacancies on boron (2.6$\div$3.2~$\%$) and yttrium (4$\div$5.5~$\%$) sites may be considered as a measure of this non-equilibrium state in YB$_6$. The increase of isolated vacancy concentration causes a structural instability development which leads to an enhancement of the electron-phonon interaction and to an $T_c$ increase in this superconductor. On the other hand, it was shown that the lowering of $T_c$ in the cage-glass structure of YB$_6$ may be attributed to the accumulation of these defects into complexes which lead to a formation of paramagnetic centers contributing to suppression of superconductivity.

Moreover, from comprehensive and detailed studies of the superconducting and normal state properties we have determined a set of parameters including the electron-phonon interaction constant $\lambda_{e-ph}$ = 0.93$\div$0.96, the critical ($H_{c1}$ and $H_{c2}$) and thermodynamic ($H_{cm}$) magnetic fields, the coherence length $\xi$(0) $\sim$ 340~${\textmd{\AA}}$, the penetration depth $\lambda$(0) $\sim$ 1100$\div$1600~${\textmd{\AA}}$ and the mean free path $l$ = 31$\div$58~${\textmd{\AA}}$, the Ginzburg-Landau-Maki parameters $\kappa_{1,2}$(0) $\sim$ 3.3$\div$4.8, the superconducting gap $\Delta$(0) $\sim$ 10.3$\div$14.8~K and the ratio 2$\Delta$(0)/$k_B T_c$ $\sim$ 4. This set of parameters points in favor of type II superconductivity in the dirty limit $\xi \gg l$  with a medium to strong electron-phonon interaction and \textit{s}-type pairing of the charge carriers in YB$_6$.

\section*{ACKNOWLEDGMENTS}

We would like to thank G.~E.~Grechnev, P.~Samuely, S.~Gab\'{a}ni and V.~Moshchalkov for helpful
discussions. The study was supported by RFBR Project No. 15$-$02$-$02553a. The measurements
were partially carried out in the Shared Facility Centre of the
P.~N.~Lebedev Physical Institute of RAS. K.F. acknowledges partial support by Slovak
agencies VEGA (2/0032/16) and APVV (14$-$0605).

%%%% ----------------------------------------------------------------------

\end{document}